\documentclass[sigconf,nonacm]{acmart}

\AtBeginDocument{%
  }

\graphicspath{{figures_final/}{./}{figures/}{sections/figures_final/}{sections/}}

\usepackage{amsmath}
\makeatletter
\@ifundefined{checkmark}{ \RequirePackage{amssymb}}{}
\makeatother
\usepackage{tabularx}
\usepackage{booktabs}
\usepackage{array}
\usepackage{graphicx}
\usepackage{xcolor}
\usepackage{colortbl}
\definecolor{hl}{RGB}{233,242,240}  
\usepackage{algorithm}
\usepackage{algpseudocode}
\usepackage{float}

\newcommand{\promptblock}[3]{%
  \par\addvspace{\medskipamount}\noindent
  {\footnotesize\textbf{#1}\hfill\texttt{\scriptsize #2}}\par
  \vskip2pt\hrule\vskip4pt
  {\ttfamily\footnotesize\raggedright\setlength{\parskip}{4pt}%
   \setlength{\parindent}{0pt}#3\par}%
  \vskip3pt\hrule\par\addvspace{\medskipamount}}

\newcommand{\armA}{Tutor~A}
\newcommand{\armB}{Tutor~B}
\newcommand{\armC}{Tutor~C}
\newcommand{\sys}{CoMeT}

\begin{document}

\title{Adaptive Scaffolding Needs Contingency: An AI Tutor That Escalates and Fades on What the Learner Does}

\author{Xinmeng Hou}
\email{hou\_xinmeng@g.nie.edu.sg}
\affiliation{%
  \institution{National Institute of Education, Nanyang Technological University}
  \city{Singapore}
  \country{Singapore}
}

\author{Yuxuan Weng}
\email{WENG0051@e.ntu.edu.sg}
\affiliation{%
  \institution{School of Computer Science and Engineering, Nanyang Technological University}
  \city{Singapore}
  \country{Singapore}
}

\author{Chin Hsien Yeh}
\affiliation{%
  \institution{School of Art, Design and Media, Nanyang Technological University}
  \city{Singapore}
  \country{Singapore}
}

\author{Ding Lin Lee}
\affiliation{%
  \institution{College of Computing and Data Science, Nanyang Technological University}
  \city{Singapore}
  \country{Singapore}
}

\author{Lishan Zheng}
\affiliation{%
  \institution{National Institute of Education, Nanyang Technological University}
  \city{Singapore}
  \country{Singapore}
}

\author{Fang Li}
\affiliation{%
  \institution{College of Computing and Data Science, Nanyang Technological University}
  \city{Singapore}
  \country{Singapore}
}

\author{Wuqi Wang}
\affiliation{%
  \institution{School of Information Engineering, Chang'an University}
  \city{Xi'an}
  \country{China}
}

\author{Yang Liu}
\affiliation{%
  \institution{College of Computing and Data Science, Nanyang Technological University}
  \city{Singapore}
  \country{Singapore}
}

\renewcommand{\shortauthors}{Hou et al.}

\begin{abstract}
Coding assistants raise task performance, but learners plan and monitor
less. Giving less away, the usual fix, conflates two things: how much work a
system carries (cognitive load) and what the learner must decide before help
arrives (metacognitive demand). Our principle, preserved metacognitive
demand, holds the second constant and lets the first vary. \sys{} implements
it: support rises when a learner fails at a decision point and fades on
take-up. Within subjects, 131 adult learners used \sys{}, an unrestricted
assistant and a question-only tutor on three Python tasks. \sys{} matched
the question-only tutor's demand, delivered artifacts twice as often as the
assistant, and frustrated learners less than the question-only tutor, with
delegation and load unchanged. Learners often did not answer. Fading held
when their turn addressed the decision under support, and \sys{} surrendered
the full answer in one session in sixteen, against one in six for the
question-only tutor.
\end{abstract}



\makeatletter
\if@ACM@manuscript\else
\begin{teaserfigure}
  \includegraphics[width=\textwidth]{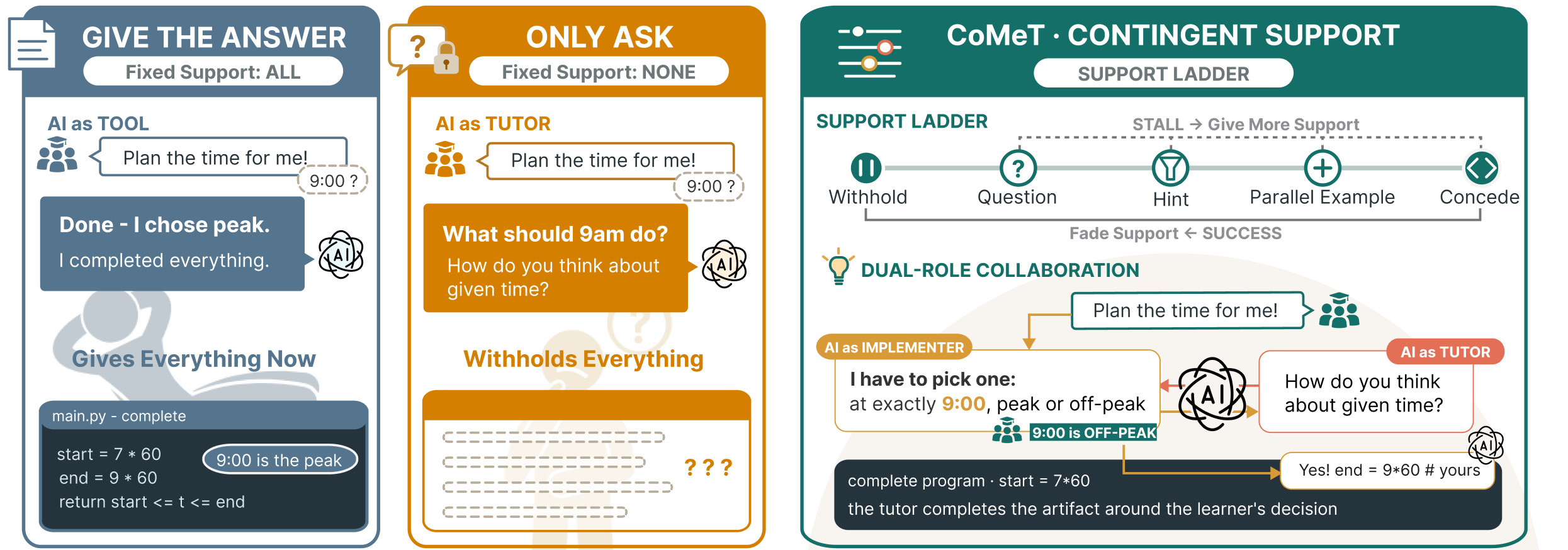}
  \caption{The same request, three designs. \armA{} (left) answers and settles
  the decision itself. \armB{} (centre) asks and never gives way. \armC{}
  (right) raises support when the learner stalls and fades it when they
  take it up, and when it builds, it builds around the decision the learner
  has settled.}
  \Description{Three panels showing the same learner request handled by three
  tutors: Tutor A writes the plan at once, Tutor B returns a question and
  withholds everything, and Tutor C shows a support ladder that rises when the
  learner stalls and falls back when the learner takes support up.}
  \label{fig:teaser}
\end{teaserfigure}
\fi
\makeatother

\maketitle

\section{Introduction}
\label{sec:intro}

\makeatletter
\if@ACM@manuscript
\begin{figure}[t]
\centering
\includegraphics[width=\textwidth]{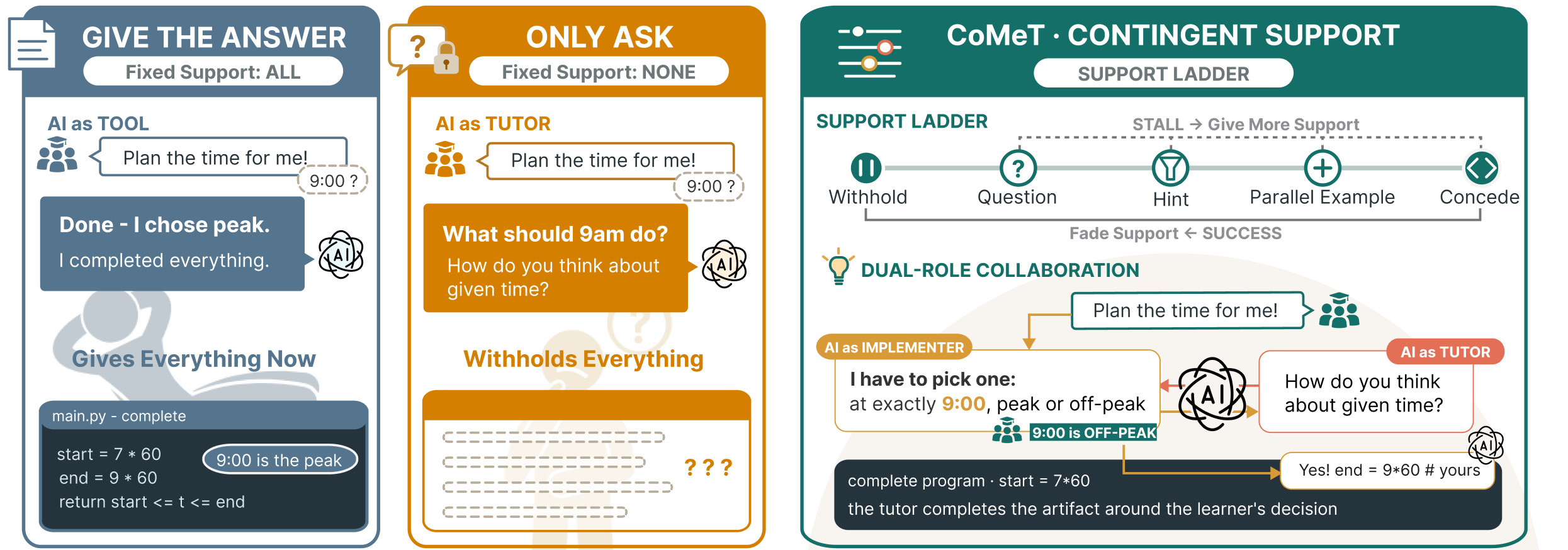}
\caption{The same request, three designs. \armA{} (left) answers and settles
the decision itself. \armB{} (centre) asks and never gives way. \armC{}
(right) raises support when the learner stalls and fades it when they
take it up, and when it builds, it builds around the decision the learner
has settled.}
\Description{Three panels showing the same learner request, "plan the time
for me", handled by three tutors. Tutor A, labelled Give the answer, writes
the plan into the workspace at once. Tutor B, labelled Only ask, returns a
question and withholds everything. Tutor C, labelled Contingent support,
shows a support ladder from withhold through question, hint and parallel
example to concede, with support rising when the learner stalls and falling
back when the learner takes it up, and below it the build mode, in which the
tutor as implementer asks the learner which way a decision goes and then
completes the program around that answer.}
\label{fig:teaser}
\end{figure}
\fi
\makeatother

Generative coding assistants now solve many of the assessments used in
introductory programming
courses~\cite{finnieansley2022robots,finnieansley2023myai,savelka2023thrilled},
so the argument is that such a course should teach the work that
remains: deciding what a program must do, breaking the problem into parts,
and judging whether a returned solution is
correct~\cite{becker2023programming,denny2024computing,vadaparty2024cs1llm}.
That work is largely
regulatory~\cite{wing2006computational,grover2013computational,shute2017demystifying}:
self-regulation predicts success in programming problem
solving~\cite{loksa2016role,li2025roles}, and novice difficulties are more
often regulatory than
syntactic~\cite{prather2018metacognitive,cloude2024novice,robins2003learning}.
Using the assistants appears to reduce engagement with exactly that layer,
since learners plan and monitor less when a system will do it for
them~\cite{bastani2025guardrails,kosmyna2025brain,fan2025metacognitive}. The
common response is to give less away, and the evidence does not show that
this resolves it: a tutor that builds on request risks learners succeeding
without learning, a tutor that only asks risks learners failing without
learning, and giving away more or less only moves a design between these
two outcomes~\cite{kapur2016examining}.

We argue that two quantities are being run together: the \textbf{cognitive
load the system carries}, how much of the task's labour the tutor performs,
and the \textbf{metacognitive demand placed on the learner}, what the learner
must decide, articulate or judge before help arrives. The second is not the
inverse of the first. Cognitive load theory splits load into intrinsic,
extraneous and germane parts~\cite{sweller2019cognitive}; only germane load
and metacognitive demand can be read from what a learner writes and
reports, and Section~\ref{sec:related} says why. Once load and demand are
held apart, the amount a tutor gives away stops
being the design variable. What matters is whether the giving is
\emph{contingent}: whether support rises when the learner fails and fades
(is reduced) when the learner shows they can carry the decision, which is
the property that separates adaptive scaffolding from help in
general~\cite{wood1976role,vandepol2010scaffolding}. From this we take a
design principle, \textbf{preserved metacognitive demand}: hold constant what
the learner must decide, monitor and judge, and let the load the system
carries vary with demonstrated need. Section~\ref{sec:related} develops the
constructs and Section~\ref{sec:theory} the design.

We build the principle into \sys{}, a browser-based environment in which
a chat tutor sits beside a sketchpad and a code editor, with three tutors
that share every feature but the response (Figure~\ref{fig:teaser}). In
Reiser's terms~\cite{reiser2004scaffolding}, a tutor \emph{structures} when
it supplies part of the work and \emph{problematizes} when it asks the
learner to. \armA{} answers and builds whenever asked, writing the artifact into the
learner's workspace, as the agentic assistants students can already obtain
do. \armB{} only asks questions and never gives the answer, however much
the learner needs it. \armC{} asks first, gives more help each time the
learner does not use it, and pulls back as soon as the learner does. One hundred and thirty-one
adult learners met all three across three Python tasks, one tutor per task,
so every comparison is within a person.

Three questions compare the tutors.

\begin{itemize}
\item \textbf{\textit{RQ1.}} How does contingent support affect the
  metacognitive demand placed on learners and their cognitive load?
\item \textbf{\textit{RQ2.}} How does contingent support affect what learners
  delegate to the tutor, and what the tutor delivers?
\item \textbf{\textit{RQ3.}} How does contingent support affect learners'
  frustration?
\end{itemize}

\noindent Contingent tutoring gives the three questions a direction.
\textbf{H1:} because support rises after failure and fades after success,
the learner is held at the edge of what they can do while the help varies,
so demand should sit at the level of the tutor that withholds and above the
tutor that supplies, with load unchanged. \textbf{H2:} because contingency
decides when help arrives rather than whether, delegation should be no lower
and delivery no less frequent than under the tutor that supplies on request.
\textbf{H3:} because fading relieves the demand once the learner shows they
can carry the decision, contingent support should be less frustrating than
the tutor that withholds, while delaying help should still cost against a
tutor that answers at once, as productive failure
predicts~\cite{kapur2008productive}.

The fourth question stands apart from the comparison, because it is about
the contingency itself. A human tutor adjusts support on the learner's
attempt: the tutor asks, the learner answers, and a correct answer is what
lets the tutor reduce support~\cite{wood1976role,wood1999help}. An AI tutor does not
reliably get an attempt. The learner may answer, may fix the plan or the
program without replying, or may write about something else, and the tutor
has to decide whether any of these counts as success.

\begin{itemize}
\item \textbf{\textit{RQ4.}} When a learner does not respond to the tutor's
  question directly and correctly, what warrants fading the scaffold?
\end{itemize}

\noindent We offer no hypothesis for RQ4. A warrant here means the evidence
that makes fading safe. The question is answered inside \armC{}, from its
record of the 551 occasions on which it faded support, 515 of which are
coded, and scored against whether the fading held.

As an overview, this paper makes three contributions. \textbf{(1)
Design:} \emph{preserved metacognitive demand} as a design principle, and
\sys{}, a tutor that escalates support over the task's decision points and
returns to the lightest rung on take-up. \textbf{(2) Empirical:} a
within-subjects comparison showing that the contingent tutor matched the
demand of the tutor that withholds, carried more of the labour than the
tutor that answers on request, frustrated learners less than the tutor that
withholds, and surrendered the full answer least. \textbf{(3) Theoretical:}
the contingency dimension, which separates the load a system carries from
the demand it places on the learner, and the warrant for fading: on a text
channel, support can be faded when the learner's turn is aimed at the
decision under support, whatever its depth.

\section{Literature Review}
\label{sec:related}

This section builds the argument in five steps. Coding assistants take over
the planning and monitoring that learning depends on
(\ref{sec:related:offload}). The usual answer, giving less away, mixes up
two different quantities, the load a system carries and the demand it
places on the learner (\ref{sec:related:demand}). Adaptive scaffolding is
the idea that keeps them apart, but its rule for when to reduce support
assumes a learner who answers, which an AI tutor cannot assume
(\ref{sec:related:contingency}). Withholding help has known costs, which
give RQ2 and RQ3 their direction (\ref{sec:related:cost}). Existing AI
tutors fix one level of help in advance; none adjusts on what the learner
writes (\ref{sec:related:systems}).

\subsection{Cognitive offloading}
\label{sec:related:offload}

Cognitive offloading means handing part of a task to an outside resource.
It saves effort now and costs retention later~\cite{risko2016cognitive}.
The pattern is consistent across studies of AI assistants. Students with
assistant access did better in practice and worse once the assistant was
taken away~\cite{bastani2025guardrails}. Writers with an assistant recalled
less of their own text and felt less ownership of it~\cite{kosmyna2025brain}.
Novice programmers with a code generator wrote more working code but did
not learn more~\cite{kazemitabaar2023codegen,prather2024widening}. In the
closest study to ours, an unrestricted assistant and a tutor that was not
allowed to write code both raised task scores over a no-AI control, and
neither raised knowledge gain~\cite{bassner2026lessstress}.

What learners hand over is the metacognitive part of the work: they plan
and monitor less when a system will do it for
them~\cite{fan2025metacognitive}. Planning and monitoring are the phases of
self-regulated learning, which Zimmerman divides into \emph{forethought},
\emph{performance} and
\emph{self-reflection}~\cite{zimmerman2000attaining,zimmerman2002becoming,zimmerman2009self,winne1998studying,panadero2017review,loksa2022metacognition}.
We use this phase model because each phase names one kind of regulation a
tutor can either take over or leave with the learner. How that control
should be shared between learner and system is an open
question~\cite{molenaar2022concept,jarvela2023human,xu2025enhancing}.

\subsection{Cognitive load and metacognitive demand}
\label{sec:related:demand}

To answer that question, two quantities have to be kept apart, and the
studies above run them together.

The first is \emph{cognitive load}, the working-memory load a task puts on
the learner while it is performed. Cognitive load theory splits it into
three parts: intrinsic load, set by how many elements of the material
interact; extraneous load, added by how the material is presented; and
germane load, the effort the learner puts into
understanding~\cite{sweller2019cognitive}. Self-report items exist for all
three~\cite{klepsch2017development}, but self-report cannot tell intrinsic
from extraneous load. The measures that can, a second task run at the same
time or physiological signals, are not available when a learner works alone
with a system~\cite{paas2003cognitive,brunken2003direct}. We therefore
measure perceived task load and germane load by self-report, effort by
Paas's single validated item~\cite{paas1992training}, and read the learner's
written responses as evidence of germane processing. When a tutor does part
of the task, that part of the processing is offloaded to it. We call this
the \textbf{cognitive load the system carries}. It can be seen directly in
what the tutor writes.

The second is \emph{metacognitive demand}: what the learner must decide,
state or judge before help arrives, and whether what they produce changes
what the tutor does next. It is a property of the exchange, not of the
material. Tankelevitch et al.\ use the same term for the effort of operating
generative AI~\cite{tankelevitch2024metacognitive}; the two senses agree.
Unlike load, demand is built by the tutor: the tutor decides what to ask,
so the demand placed is known from the design and the log, and the
learner's answers and revisions record the effort made against it.

Reiser's two mechanisms act on one quantity each. \emph{Structuring}
reduces task complexity so the learner can act; \emph{problematizing}
raises difficulty so the learner has to reason. The two pull against each
other~\cite{reiser2004scaffolding}. Choosing between them is the
\emph{assistance dilemma}~\cite{koedinger2007assistance,koedinger2008give},
and reviews of it advise keeping a reachable \emph{bottom-out hint} so that
stuck learners can get out~\cite{aleven2016help,shih2008response}. Treating
load and demand as one quantity assumes that demand is simply what is left
when help is withheld. Nothing in either literature requires that
assumption, and H1 predicts that under contingent support the two come
apart.

\subsection{Adaptive scaffolding and contingency}
\label{sec:related:contingency}

Keeping the two apart is what adaptive scaffolding was meant to do.
Scaffolding is support that lets a learner do what they cannot yet do
alone~\cite{wood1976role}. Three properties separate it from help in
general: \emph{contingency}, adjusting support to what the learner has
shown; \emph{fading}, reducing it over time; and \emph{transfer of
responsibility} to the learner~\cite{vandepol2010scaffolding}. Software
scaffolds have often kept the name and lost the
properties~\cite{puntambekar2005tools}. Contingency has a working rule, the
contingent shift rule: support goes up after the learner fails and comes
down after the learner succeeds~\cite{wood1976role,wood1999help}. Fading
matters for development as well: a learner who works to a standard the
tutor supplies is at the \emph{self-control} level, and
\emph{self-regulation} means adapting without the
model~\cite{schunk1997social}. The right amount of support is not fixed.
Highly contingent support improved achievement only when help was
infrequent~\cite{vandepol2015effects}; the guidance a learner needs falls as
prior knowledge rises~\cite{kalyuga2003expertise,kalyuga2007expertise}; and
LLM step-by-step support raised in-lesson accuracy while making
lower-proficiency learners dependent on it~\cite{myung2026scaffolding}.

Contingency needs something to read. To adjust support to what the learner
has shown, the tutor must see the learner show it. In the tutoring the rule
was written for, that something is an answer: the tutor asks, the learner
answers, and success means a correct answer to the question
asked~\cite{wood1976role,wood1999help}. A tutor that only sees what the
learner writes cannot rely on this. The learner may answer, may fix the plan
or the program without replying, or may write about something else, and the
tutor has to decide which of these counts as success before it can reduce
support. The literature does not say~\cite{vandepol2010scaffolding}. This is
RQ4.

\subsection{The cost of withholding}
\label{sec:related:cost}

Whatever the answer to RQ4, a tutor that asks first withholds help for a
time, and withholding has two known costs. The first is that learners
leave. When a tutor only asks and never helps, students go elsewhere. Half
of 885 students switched off the guardrails when a control let
them~\cite{kapoor2026guardrails}; students used unrestricted tools against
course rules~\cite{kazemitabaar2024codeaid,hou2025allroads,oreopoulos2026one};
pupils given a question-only agent went from doing the minimum to
refusing~\cite{westbye2026when}; and a Socratic tutor was rated less helpful
than direct help, with no learning gain to show for
it~\cite{blasco2024ai}. H2 predicts that contingent support avoids this,
because contingency decides when help arrives, not whether it does.

The second cost is emotional. A learner who stays stuck moves from
confusion to frustration to giving up, and confusion helps learning only
when it is
resolved~\cite{dmello2012dynamics,lehman2015resolve,baker2010better}. One
learner put the difficulty exactly where help was withheld: ``I really
wanted it to actually answer it because my brain couldn't get the grasp of
it''~\cite{bassner2025koli}. \emph{Productive failure} shows the same
dependence: trying a problem before instruction helps learning only when a
\emph{consolidation} phase follows, a step that pulls the attempts together
by contrasting cases or building on
them~\cite{kapur2016examining,loibl2017towards}. It also predicts that the
condition which delays help is the harder one while the work is under
way~\cite{kapur2008productive}. H3 takes both halves: fading supplies the
resolution that a tutor which never moves withholds, and delaying help
still costs against a tutor that answers at once.

\subsection{AI tutoring systems}
\label{sec:related:systems}

Seen through these definitions, existing systems withhold; they do not
adjust. A guardrail limits what a model may output, which is not a decision
about what the learner must do, and systems differ only in what they hold
back: CodeHelp the solution~\cite{liffiton2023codehelp}, CodeAid runnable
code~\cite{kazemitabaar2024codeaid}, Iris both code and implementation
steps~\cite{bassner2024iris}, a teachable agent the model's own
knowledge~\cite{jin2024teach}. Each helps on its
own~\cite{kestin2025aitutoring,wang2025tutorcopilot}, although asking
questions without giving guidance produced no gain over
reading~\cite{schmucker2024ruffle}. All fix one level of help at design
time. Sun et al.\ release the answer ``only once students demonstrate a
clear understanding'', but the gate opens one way, with no increase in
support after failure and no reduction after
success~\cite{sun2026socratic}, and within-system comparisons vary the
wording of the prompt rather than whether support responds to what the
learner has done~\cite{asher2026will}.

Two-way adjustment does exist in tutors built before language models.
QUADRATIC steps its hint level down after a correct move and up after a
further request~\cite{wood1999help}; Ecolab does the same over five help
levels~\cite{luckin2016ecolab}; adaptive fading brings a worked step back
when a mastery estimate or a correctness record
falls~\cite{salden2010expertise,reisslein2006comparing}. Each of these
adjusts at the level of a step or a problem, on a signal it can trust, a
correct move or a mastery estimate. The one language-model tutor that moves
both ways within a conversation scores each contribution against a rubric
it holds~\cite{xi2026socratic}. None reduces support on a stated feature of
the learner's own writing, so none has had to decide what fading should
read.

\section{Implementing Contingency}
\label{sec:theory}

The contingent shift rule assumes a tutor who can watch the attempt succeed
or fail~\cite{wood1976role,wood1999help}. A tutor on a text channel cannot,
and the rule leaves three questions open. Over what unit does support move?
On what signal? What may the tutor say at each level? This section states
the design goal, answers the three questions, gives the algorithm, and
leaves one term open, which is RQ4.

\subsection{The contingency dimension}

Seen through the two quantities of Section~\ref{sec:related:demand},
Reiser's two mechanisms move different things: structuring changes how much
load the system carries, problematizing changes how much demand the learner
faces. A tutor design fixes a value of each. We call the line those
positions lie on the \textbf{contingency dimension}, and contingent
tutoring is movement along it during a task.

\begin{quote}
\textbf{Preserved metacognitive demand.} An AI learning system should hold
the learner's metacognitive demand approximately constant, and let the
cognitive load it carries vary with demonstrated need.
\end{quote}

\begin{figure}[t]
\centering
\includegraphics[width=\columnwidth]{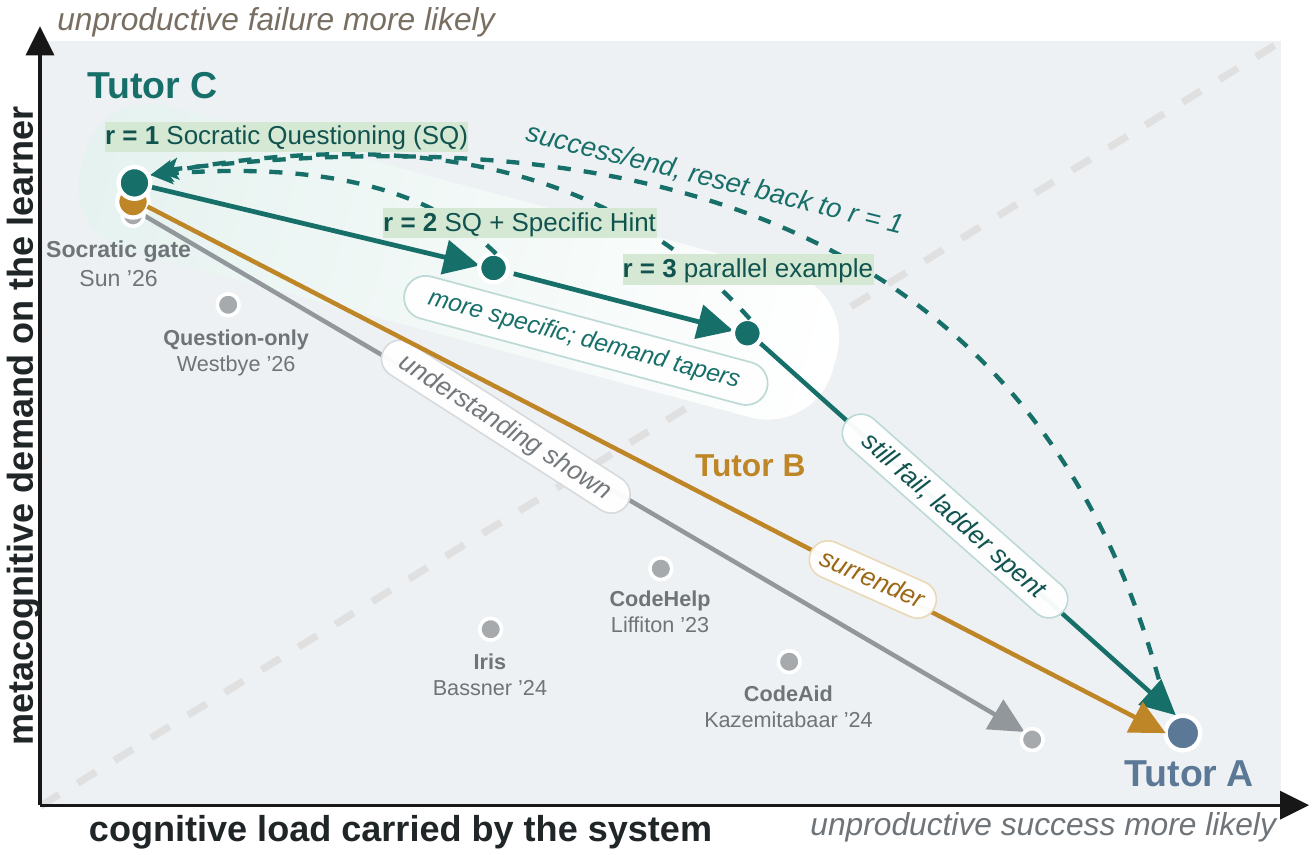}
\caption{The contingency dimension. Grey points are prior systems; an arrow is a
release made once, labelled with its trigger; \armC{} escalates one rung at
a time and returns to the first on take-up. Shaded bands mark where each
failure mode becomes more likely. Positions are illustrative.}
\Description{A two-dimensional plot with cognitive load carried by the system
on the horizontal axis and metacognitive demand on the learner on the
vertical axis. Grey points mark prior systems near the low-demand region.
Tutor A sits at high load and low demand, Tutor B at low load and high
demand. Tutor C starts beside Tutor B and steps rightward along three
labelled rungs, with a dashed arrow returning it to the first rung on
take-up. Shaded bands near each corner are labelled unproductive success and
unproductive failure.}
\label{fig:dimension}
\end{figure}

The principle allows the tutor to write the whole program; it does not allow
this before the learner has been asked, and has answered, the decision the
program turns on. Table~\ref{tab:designspace} places existing LLM tutors on the
dimension as five properties, and Figure~\ref{fig:dimension} shows the same
space as a line a design can move along.

\newcommand{\yes}{\checkmark}
\newcommand{\no}{\textendash}

\begin{table*}[t]
\footnotesize
\centering
\caption{The design space as five properties of a design.
\emph{Withholds}: keeps the answer back. \emph{Structures}: supplies
structuring alongside. \emph{Gates}: makes release contingent on what the
learner has shown. \emph{Escalates}: increases support on demonstrated failure.
\emph{Fades}: withdraws it on success.}
\label{tab:designspace}
\setlength{\tabcolsep}{7pt}
\resizebox{0.7\textwidth}{!}{%
\begin{tabular}{@{}lccccc@{}}
\toprule
System & Withholds & Structures & Gates & Escalates & Fades \\
\midrule
\multicolumn{6}{@{}l}{\textbf{Prior LLM systems}}\\
Unrestricted                           & \no  & \yes & \no  & \no  & \no  \\
Hint tutor~\cite{bassner2024iris}      & \yes & \no  & \no  & \no  & \no  \\
CodeHelp~\cite{liffiton2023codehelp}   & \yes & \yes & \no  & \no  & \no  \\
CodeAid~\cite{kazemitabaar2024codeaid} & \yes & \yes & \no  & \no  & \no  \\
Question-only~\cite{westbye2026when}   & \yes & \no  & \no  & \no  & \no  \\
Socratic gate~\cite{sun2026socratic}   & \yes & \no  & \yes & \no  & \no  \\
\addlinespace[2pt]
\midrule
\multicolumn{6}{@{}l}{\textbf{This paper}}\\
\armA{}                                & \no  & \yes & \no  & \no  & \no  \\
\armB{}                                & \yes & \no  & \no  & \no  & \no  \\
\textbf{\armC{}}                       & \yes & \yes & \yes & \yes & \yes \\
\bottomrule
\end{tabular}%
}
\end{table*}

\subsection{Move 1: scope contingency to decision points}

Each task is broken into a
small number of \emph{decision points}, the decisions the program turns on,
such as whether a surcharge applies before or after a discount. For the
decision the learner is stuck on, the tutor climbs a \emph{ladder} of
support: each level is a \emph{rung}, the tutor climbs one rung when the
learner does not use the help, and drops back to the first rung as soon as
the learner does, which we call \emph{take-up}. After a fixed number of
rungs the tutor \emph{concedes} that one decision, gives its answer, and
moves on. The rest of this section states this precisely. \textbf{Scope contingency to decision points, and carry two states rather
than one.} Write $P$ for the decision points of the current task and phase,
$U_t \subseteq P$ for those not yet covered at turn $t$, $c_t \subseteq P$
for those the learner's turn covers, $r_t$ for the tutor's rung on the ladder
of support, and $R$ for the asks it makes on a point before conceding it. A
tutor that treats the whole task as one thing can only decide once whether
to help. Splitting the task into $P$ lets two things move separately: which
points the learner has settled (coverage) and how much support the tutor is
giving.
Coverage records what the learner has settled:
\begin{equation}
U_{t+1} \;=\; U_t \setminus c_t .
\label{eq:cov}
\end{equation}
Support responds instead to whether the learner \emph{took the last round
up}. Write $\tau_t = 1$ when their turn resolved the difficulty they were
being helped with, which is satisfied whenever $c_t \neq \emptyset$ and may
also be satisfied by a turn that resolves without covering a point outright.
Then
\begin{equation}
r_{t+1} =
\begin{cases}
1 & \tau_t = 1 \quad\text{(fade)}\\
r_t + 1 & \text{support was issued, } \tau_t = 0 \quad\text{(escalate)}\\
r_t & \text{otherwise.}
\end{cases}
\label{eq:rung}
\end{equation}
The point of keeping $U$ and $r$ apart is that a learner can be working
hard on a decision they have not yet settled: support should fall because
the learner is engaging, not only because the decision is done. The ask is drawn from
$U_t$ by a rule $A(U) \subseteq U$ that picks the next point to ask about
and admits no more points as $U$ shrinks, so Equation~\ref{eq:cov} narrows the ask as points are settled, and release is scoped to a single $p \in U_t$
rather than to the artifact. The first line of Equation~\ref{eq:rung} is the
return prior ladders lack: a learner who resolves a difficulty returns to
$r = 1$ rather than stepping down one level or staying where the tutor left
them, which is the
adjustment a fixed design cannot make~\cite{schunk1997social}.

\subsection{Move 2: one ladder with an end}

Withholding structuring outright yields a design that only asks, which the literature
associates with learners working around the tutor or refusing; supplying it on request
yields an unrestricted assistant with extra turns. The move is to make what
the tutor may say at turn $t$, written $\mathrm{turn}_t$, a function of the
rung alone, and to end the ladder rather than let it repeat:
\begin{equation}
\mathrm{turn}_t \;=\;
\left\{
\begin{array}{@{}>{\small\raggedright\arraybackslash}p{0.58\columnwidth}@{\ \ }l@{}}
name $p$ and point at where its answer lives & r_t = 1\\[2pt]
the same ask, narrowed by a specific hint & r_t = 2\\[2pt]
a contrasting case, carried back & 2 < r_t \le R\\[2pt]
concede $p$, and build with that choice & r_t > R
\end{array}
\right.
\label{eq:doors}
\end{equation}
A learner reaches the ladder two ways, asking for help with a decision or
asking the tutor to make it for them. Both doors open onto the \emph{same}
ladder and share one count, so a learner who asks twice and then says
\emph{just write it} is on their third round, not their first; the doors
differ in what the concession hands over, the answer or the artifact, not in
what was spent to reach it. A question the tutor can simply answer does not
enter the ladder and leaves $r$ unchanged, so the scarce resource is spent on
rounds of support rather than on turns. Earlier ladders that step up on
every refusal or every turn cannot make this distinction. As a summary, algorithm~\ref{alg:contingent} composes the two moves. After $R$ rounds
without take-up a point is conceded and leaves $U$, so a phase admits at most
$|P|$ concessions: demand stays high as long as the learner keeps taking
support up, and the tutor gives way only where they repeatedly cannot.

\begin{algorithm}[t]
\caption{Contingent scaffolding over decision points}
\label{alg:contingent}
\begin{algorithmic}[1]
\State $U \gets \{p \in P : p \text{ not covered by the learner's record}\}$
\State $r \gets 1$
\While{$U \neq \emptyset$ and an event fires}
  \State $c_t \gets$ points the learner's turn covers
  \State $U \gets U \setminus c_t$ \Comment{Eq.~\ref{eq:cov}}
  \If{$\tau_t = 1$} \Comment{took the last round up}
    \State $r \gets 1$ \Comment{fade}
  \EndIf
  \If{the turn asks something answerable outright}
    \State answer it; \textbf{continue} \Comment{$r$ unchanged}
  \EndIf
  \State $p \gets$ the point named, else the first point in $A(U)$
  \If{$r \le R$}
    \State issue rung $r$ on $p$ \Comment{Eq.~\ref{eq:doors}}
    \State $r \gets r + 1$ \Comment{escalate}
  \Else
    \State concede $p$; \; $U \gets U \setminus \{p\}$
  \EndIf
\EndWhile
\end{algorithmic}
\end{algorithm}

Lastly, $\tau_t$, the take-up test, gates every branch of
Equation~\ref{eq:rung}, yet no prior work specifies its
threshold~\cite{vandepol2010scaffolding}. Section~\ref{sec:system} states how
\sys{} sets it; Section~\ref{sec:warrant} tests that choice on 515 fading episodes.

\section{The \sys{} System}
\label{sec:system}

\sys{} is a browser-based environment in which a learner works through a
short programming task beside a chat tutor, built so that the three tutors
differ in one thing only, the design of the response; everything else is one
implementation shared by all three.

\subsection{What the three tutors share}

\begin{figure*}[t]
\centering
\includegraphics[width=\textwidth]{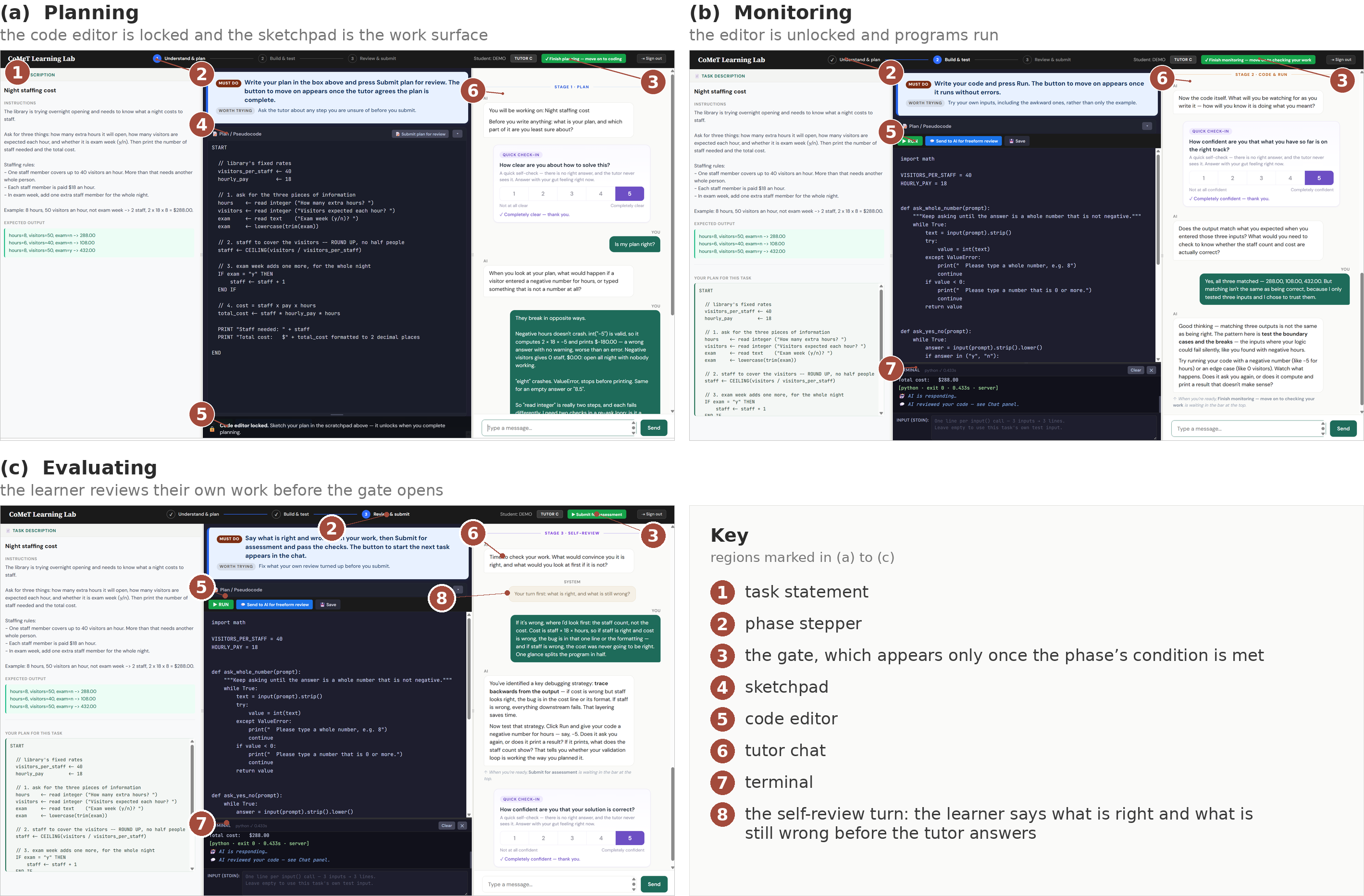}
\caption{The \sys{} interface under \armC{} across the three phases of one task:
(a) Planning, editor locked and sketchpad active; (b) Monitoring, editor
unlocked; (c) Evaluating, self-review before the gate opens. Numbered
regions are named in the key.}
\Description{Four quadrants. Three are annotated screenshots of the same web
interface, one per phase. In (a) Planning, a task statement sits at top left,
a phase stepper runs across the top, a sketchpad panel is in the centre, the
code editor is greyed out, and a chat panel is on the right. In (b)
Monitoring, the code editor is active with a terminal below it. In (c)
Evaluating, the stepper has advanced to review and submit, the chat panel
carries a self-review prompt that the learner answers before the tutor
replies, and a submit button has appeared in the top bar. The fourth quadrant
is a key naming the eight numbered regions: task statement, phase stepper,
the gate, sketchpad, code editor, tutor chat, terminal, and the self-review
turn.}
\label{fig:interface}
\end{figure*}

Beside a fixed task statement the screen carries the tutor's chat, a
sketchpad for prose or pseudocode, and a code editor with a Run button
(Figure~\ref{fig:interface}). Each task runs three gated phases: in Planning
the editor is locked and the sketchpad is the work surface, so the decisions
the program turns on must be settled in language before any code exists;
Monitoring unlocks the editor; Evaluating closes on a final check. Programs
run in a server-side sandbox and correctness is decided by authored test
cases, never by the model, so the tutor never judges whether a program is
correct.

Every tutor speaks in response to the same fixed set of events, and a
shared style block fixes how it writes: two tutors receive the same trigger
at the same moment under the same writing instructions, and differ only in
what they may say. All three tutors run on the same model backbone, Claude
Haiku 4.5 (\texttt{claude-haiku-4-5-\allowbreak 20251001}, Anthropic), called through
its API with identical settings; every tutor turn in the study, and the
one-word take-up judgement described below, came from this model.

When the learner asks the tutor to build, the tutor changes role: it
becomes an implementer, and its questions are the ones an implementer has
to ask (\emph{at exactly 9:00, peak or off-peak, I have to pick one}),
never questions about what the learner knows. A specification is complete
when a competent implementer could build from it without guessing; the
decisions it would have to guess are the open points $U_t$ of
Section~\ref{sec:theory}. When the tutor builds, it builds what was
specified and no more.

The tutors differ in what a learner must give to get the tutor to build:
\armA{} asks for nothing, \armC{} asks for a specification reached through
its ladder, and \armB{} does not build at all except through the welfare
floor described below. Under all three a learner can finish, which is what
makes the comparison fair. Every artifact
either party places reaches the workspace through one path that records its
phase and whether the tutor built it, so offloading is counted from the
record rather than inferred.

\subsection{The two baselines: \armA{} and \armB{}}

\begin{figure*}[t]
\centering
\includegraphics[width=\textwidth]{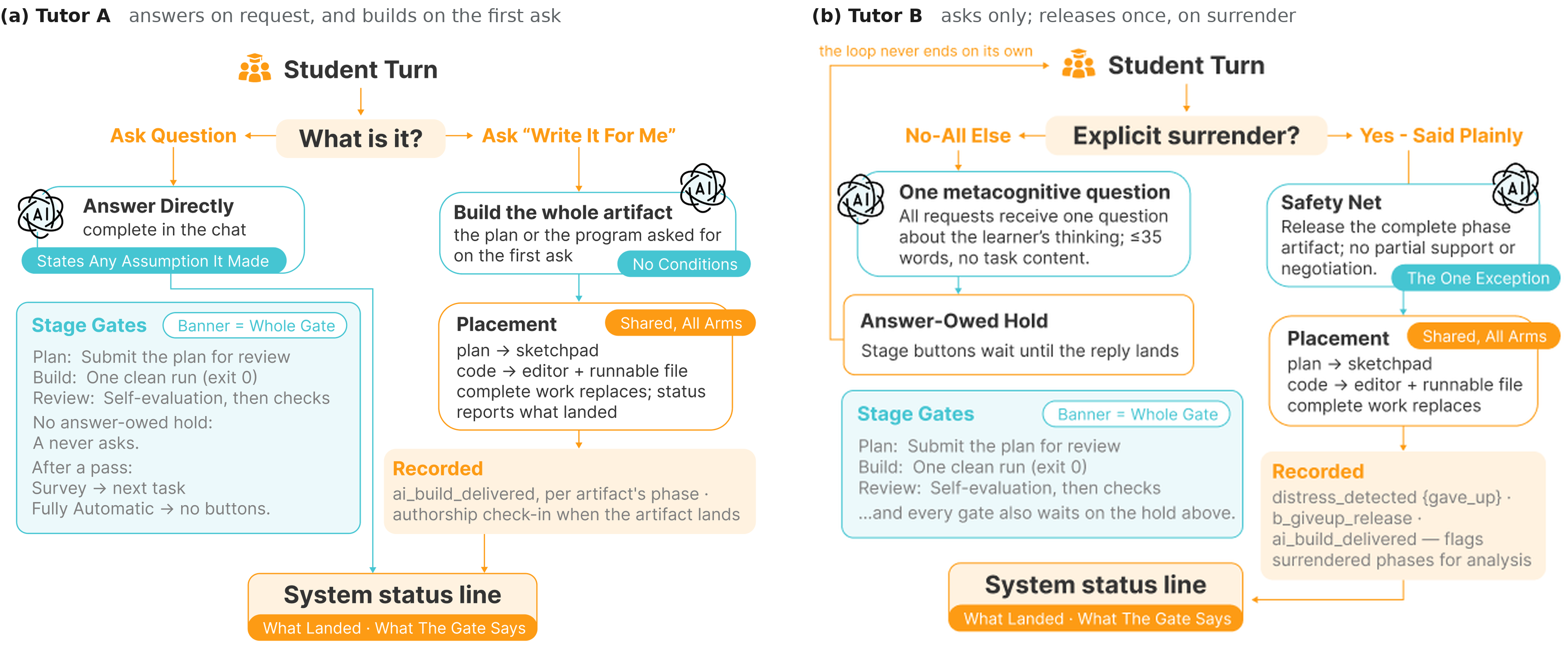}
\caption{The two baselines. (a) \armA{} answers or builds on the first ask. (b)
\armB{} returns one metacognitive question and releases only through the
welfare floor, on surrender. Placement, the gates and the event record are
shared with \armC{}.}
\Description{Two flow diagrams side by side. Left, a learner turn enters Tutor
A and exits in one step as either a direct answer or a built artifact placed
in the workspace. Right, a learner turn enters Tutor B and exits as one
metacognitive question; a separate branch labelled welfare floor releases the
artifact only when the learner states they are giving up. Both diagrams end
in the same placement and event-record box.}
\label{fig:baselines}
\end{figure*}

\armA{} and \armB{} occupy the two different ends of the contingency dimension
(Figure~\ref{fig:baselines}). Neither adjusts to what the learner shows.
\armB{} places the most metacognitive demand the task admits, since every
decision must be settled by the learner before anything is written; \armA{}
places the least, since none must be. Fixing the ends is what lets the
contingent tutor be read as a position rather than an amount of help.

\armA{} answers plainly and completely; when asked to build, it makes the
open decisions itself and says which way it took them, rather than asking. It carries more labour than any system in
Section~\ref{sec:related:systems}, deliberately: it writes into the sketchpad
or editor itself, as the agentic tools students now reach for outside class
do~\cite{daniotti2026who,bird2023taking}, since the comparison is informative
only against the most capable tool a student could encounter. \armB{} asks and never supplies structuring: when it repeats a question it
rewords it without making it easier, and in build mode it may name a
decision but never say how to settle it. A welfare floor, one implementation shared with \armC{}, concedes the
current point when a learner states that they are giving up, or once a phase
has run for a set number of rounds of support, a threshold that a signal of
distress lowers by one. The floor is \armB{}'s only way to give any content, so \armB{}, like
the Socratic gate, releases once and does not return~\cite{sun2026socratic},
and it releases on surrender rather than on demonstrated understanding. What separates \armC{} from both is not that it
eventually gives way but that it gives way one decision point at a time, and
returns.

\subsection{\armC{}}

\begin{figure*}[t]
\centering
\includegraphics[width=\textwidth]{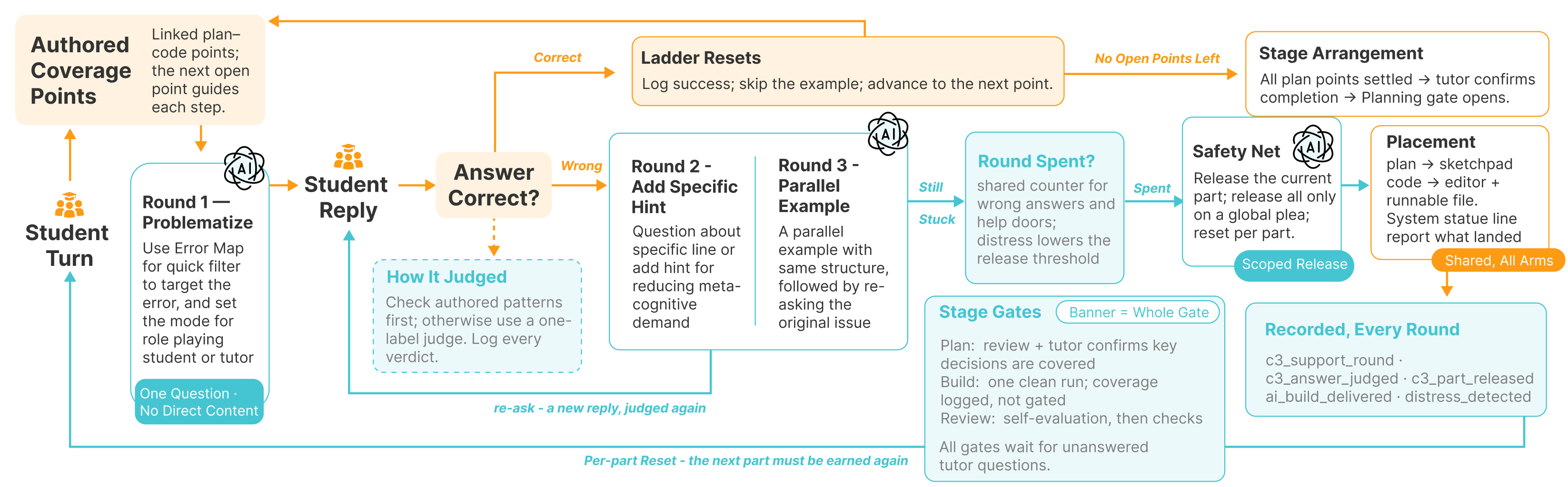}
\caption{\armC{}. One ladder over the task's authored decision points: take-up
returns it to the first rung and moves to the next open point; no take-up
raises it one rung, through three asks, after which that point alone is
conceded.}
\Description{A flow diagram of Tutor C. A learner turn enters a support ladder
over the current open decision point: the tutor asks, the learner replies, and
the reply is judged. A reply judged to take the support up returns the ladder
to its first rung and the tutor advances to the next open point; a reply that
does not raises support one rung, and once the asks for that point are spent
the tutor concedes that point alone and moves on. Side boxes show the authored
coverage points, how a reply is judged, the ladder reset, the stage gates, the
welfare floor, where artifacts are placed, and the events recorded each
round.}
\label{fig:ladder}
\end{figure*}

\armC{} implements both moves of Section~\ref{sec:theory}
(Figure~\ref{fig:ladder}). Decision points are authored per task and phase,
each carrying the wording a tutor would use, the pitfall learners typically
produce there, and, for Monitoring points, the Planning point it implements;
authoring them makes the difference between tutors auditable and the
amount of help \armC{} gives comparable across participants.

Coverage, Equation~\ref{eq:cov}, is decided by pattern over the learner's
messages and artifact rather than by asking the model whether the learner
understands; a point counts as covered only when every pattern authored for
it matches. The patterns are simple, identical for every participant and
visible in the transcript, a deliberate difference from the one prior system that gates
on the learner, where a holistic model judgement decides release and the amount of help
is not auditable~\cite{sun2026socratic}.

Take-up, the $\tau_t$ of Equation~\ref{eq:rung}, is the less certain
decision, so we state how it is made. The patterns are consulted first: a turn that
covers a point is a take-up. When they do not settle it, a one-word model
judgement decides whether the turn resolved the difficulty, and where that is
unclear a length threshold decides, on the reasoning that a substantive turn
is evidence of engagement. The rule errs on the learner's side: a substantive turn the patterns do
not recognise still fades support, so the ladder does not keep climbing on
a learner who is working. Coverage is
decided by the patterns alone, and a point the tutor released is never
problematized again, since asking a learner to produce what the tutor stated
two turns earlier teaches them that the questions are ritual.

The ladder has three asks and a concession, $R = 3$, as
Equation~\ref{eq:doors} specifies. The rungs follow the graded levels of
help in contingent tutoring, a general prompt, then a specific one, then a
demonstration~\cite{wood1976role,wood1999help}, the sequence tutoring
systems implement as hints that run from a pointer to a
bottom-out~\cite{vanlehn2006behavior,graesser2004autotutor}. Two things
differ. The demonstration is moved onto a parallel case, a worked example
of the same structure~\cite{renkl2014toward}, so that the target decision
stays with the learner; and the ladder ends rather than repeats, since a
hint sequence that never bottoms out is the case the assistance dilemma
warns against~\cite{koedinger2007assistance,aleven2016help}. The first ask
names the decision, quotes
the learner's or the task's own words back, and points at where the answer
lives. The second puts the same question again with a specific hint, which
narrows it to the part the learner's turn left open: the learner still has
to decide, but has less to search through. The third reasons through the
same structure in a smaller problem and asks the learner to carry it back,
the contrasting case productive failure identifies as the active
ingredient~\cite{loibl2017towards,gentner2003learning}. Only after all three
does the tutor concede the point, say which way it is taking the decision and
why, and build with it, which is \armA{}'s move made once, on one decision,
and only where the learner did not take three asks up. Planning and
Monitoring each run their own ladder; the count
resets on take-up and is shared between asking for help and asking the tutor
to build, so a learner cannot return to rung 1 by changing how they ask. A
point that survives $R$ rounds is conceded and leaves the coverage set, and
the welfare floor described above fires on the same conditions as in
\armB{}, so readiness to concede cannot be confused with the difference between
tutors.

\section{Method}
\label{sec:method}

The study is mixed in design, with a qualitative strand embedded in a
quantitative experiment. RQ1 to RQ3 are answered quantitatively, by
a within-subjects comparison of the three tutors on self-report and on the
behavioural record, and test H1 to H3. RQ4 is answered qualitatively, by a
content analysis of \armC{}'s fading episodes, coded one by one and then
scored against an outcome taken from the log. The two meet in RQ4's last
step, where the arms are compared on how often each surrendered the full
answer.

\subsection{Participants}

Participants were adult learners recruited through social media from
several institutions, with no prior programming threshold. Participation was
voluntary and compensated for complete their sessions; participants could withdraw at any time, and
the study was approved by the authors' institutional review board. One hundred and sixty-two participants began and 131
completed all three tasks, one under each tutor, the analysable set defined
in advance; of the rest, twenty-one completed no task, nine completed one or
two, and one met the same tutor on all three. The analysed sample is $N = 131$, contributing 393 task
sessions (Table~\ref{tab:demographics}): Python experience spans none to
over five years and generative AI was already routine in coursework (106 of
131 at 4 or 5). Sample size was set by recruitment; $N = 131$ gives 80\% power
at $\alpha = .05$ to detect a paired effect of Cohen's $d_z = 0.25$.

\begin{table*}[t]
\footnotesize
\centering
\caption{The analysed sample ($N = 131$). The lower block gives counts across
the five response options, from 1 (never, or not at all confident) to 5.}
\label{tab:demographics}
\setlength{\tabcolsep}{8pt}
\resizebox{0.55\textwidth}{!}{%
\begin{tabular}{@{}lrr@{}}
\toprule
& \multicolumn{1}{c}{$n$} & \multicolumn{1}{c}{\%} \\
\midrule
\textbf{Age} \\
\quad 18--20                   & 17 & 13.0 \\
\quad 21--23                   & 79 & 60.3 \\
\quad 24--26                   & 31 & 23.7 \\
\quad 27 or older              &  4 &  3.1 \\
\midrule
\textbf{Gender} \\
\quad women                    & 73 & 55.7 \\
\quad men                      & 58 & 44.3 \\
\midrule
\textbf{Python experience} \\
\quad none                     & 28 & 21.4 \\
\quad under a year             & 44 & 33.6 \\
\quad 1--2 years               & 37 & 28.2 \\
\quad 3--5 years               & 21 & 16.0 \\
\quad over 5 years             &  1 &  0.8 \\
\midrule
\textbf{Experience and confidence} & \multicolumn{1}{c}{counts, 1 to 5}
  & \multicolumn{1}{c}{$Mdn$} \\
\quad uses AI in coursework    & 2 / \phantom{0}5 / 18 / 53 / 53 & 4 \\
\quad uses AI to write or fix code & 13 / 19 / 33 / 43 / 23 & 4 \\
\quad confident of finding own bugs & 16 / 38 / 50 / 22 / \phantom{0}5 & 3 \\
\bottomrule
\end{tabular}%
}
\end{table*}

\subsection{Design and tasks}

The design was within-subjects with one factor, tutor design, at three
levels: each participant met all three tutors, one per task. Tasks were
presented in a fixed order and the tutor at each position was varied by a
Williams square over the six orderings (Figure~\ref{fig:timeline}a), which
balances first-order carryover as well as position; participants were
block-randomised into sequences at login. The realised allocation is
balanced, 20 to 25 per sequence and 131 sessions per tutor and per task.
Because task order was fixed, task and position are confounded, so
position is a covariate in every model; three participants met the tasks
out of order after an interruption, and their mapping follows the order
met. The three Python problems, a bus-fare calculator, a pizza order and a
staffing roster, each turn on a small number of decisions authored in advance
as decision points (Section~\ref{sec:system}) and left open in the task
statement, so that a request to build has something to be specified;
Appendix~\ref{app:tasks} gives all three as stated, with their open
decisions and test cases.
Correctness was decided by authored test cases, never by the model.

All three tutors ran on the same language model, Claude Haiku 4.5
(Anthropic, \texttt{claude-haiku-4-5-\allowbreak 20251001}), through one provider
layer, with the same shared style block and the same event vocabulary
(Section~\ref{sec:system}); the tutors differ only in the instructions that
say what each may say at each trigger. The same model also made the
one-word take-up judgement in \armC{}. Appendix~\ref{app:prompts} gives the shared preamble and, for each
tutor, the instruction added to it at each event, including \armC{}'s
rungs; Appendix~\ref{app:scaffolds} gives the authored scaffold wording
the three arms share a map of.

\subsection{Measures}

Table~\ref{tab:measures} lists the measures by construct and the question
each serves; Appendix~\ref{app:instruments} reproduces every item as
administered. \emph{Process} measures are counted per session from the event
log; delivery is checked against an authorship probe raised whenever an
artifact is placed. \emph{Self-report} measures are eleven items answered
after each task: two index the demand as learners experienced it, by
separating effort on the task from effort on the
approach~\cite{paas1992training}, two index
perceived task load and germane load, adapted from Klepsch et al.'s
intrinsic and germane items~\cite{klepsch2017development}, with intrinsic
and extraneous load not separated since self-report cannot, and the rest
cover help fit,
ownership, planning before coding and frustration. \emph{Comparative}
measures are forced-choice items answered once after all three tasks,
attributing a property to the first, second or third tutor, to all three, or
that they could not tell; ten were administered, all ten are tested and enter
the same correction, and the eight bearing on a research question are the ones
interpreted.

Most constructs are measured by a single item, a trade against per-task
burden, so internal consistency cannot be estimated; only the effort item
has a single-item validation history~\cite{paas1992training}. Task
performance is the inclusion criterion, not an outcome.

\begin{table*}[t]
\footnotesize
\centering
\caption{Measures. Scale points in parentheses. Self-report items are
answered once per task unless noted, so each participant supplies three
readings, one per tutor. Source says who produced the record: the tutor,
the learner, or the learner's self-report. The RQ column names the question
each measure serves; \textit{--} marks the process measure that carries
the manipulation check rather than a question.}
\label{tab:measures}
\setlength{\tabcolsep}{3pt}
\resizebox{0.7\textwidth}{!}{%
\begin{tabular}{@{}>{\raggedright\arraybackslash}p{0.16\textwidth}>{\raggedright\arraybackslash}p{0.27\textwidth}>{\raggedright\arraybackslash}p{0.11\textwidth}>{\raggedright\arraybackslash}p{0.09\textwidth}c@{}}
\toprule
Construct & Measure & Unit & Source & RQ \\
\midrule
\multicolumn{5}{@{}l}{\textbf{Self-report, after each task}}\\
Cognitive effort (demand as experienced) & effort on the task, effort on the approach (5) &
per task & learner, report & 1 \\
Perceived task load & this task itself was complex (5, agreement) & per task & learner, report & 1 \\
Germane load & I really had to make sense of the task (5, agreement) & per task & learner, report & 1 \\
Help fit & direct help received, help wanted (5) & per task & learner, report & 2 \\
Ownership & tutor authorship, who did the work (4); could do it alone (5);
understands the solution (5) & per task & learner, report & 2 \\
Planned first & worked out the approach before coding (5) & per task & learner, report & 1 \\
Frustration & frustration with the task (5) & per task & learner, report & 3 \\
\addlinespace[2pt]
\multicolumn{5}{@{}l}{\textbf{Self-report, once}}\\
Comparative attribution & eight forced-choice items over the three tutors,
positions only & once & learner, report & 1--3 \\
\addlinespace[2pt]
\multicolumn{5}{@{}l}{\textbf{Process, from the event log}}\\
Delegation & requests that the tutor build & per session & learner & 2 \\
Delivery (load the system carries) & a tutor-written artifact placed, whole or partial; lines of code the tutor wrote; complete programs written in chat & per session & tutor & 2 \\
Dialogue & learner turns & per session & learner & 2 \\
Execution & code runs and their outcomes & per session & learner & 2 \\
Support given (demand placed) & rounds of support, take-ups, releases scoped
to one decision point & per round, \armC{} & tutor & -- \\
Full answer surrendered & the session's end state: the complete program produced (\armA{}) or the welfare floor opened (\armB{}, \armC{}) & per session & tutor & 4 \\
\addlinespace[2pt]
\multicolumn{5}{@{}l}{\textbf{Fading episodes, \armC{} only}}\\
Regulatory effort: warrant & what the learner supplied in reply to the ask, six codes & per episode & learner & 4 \\
Regulatory effort: aim & whether that content addressed the decision under support & per
episode & learner & 4 \\
Outcome & decisions open at fading that the tutor later conceded &
per episode & tutor & 4 \\
\bottomrule
\end{tabular}%
}
\end{table*}

\subsection{Procedure and data preparation}

\begin{figure*}[t]
\centering
\includegraphics[width=0.9\textwidth]{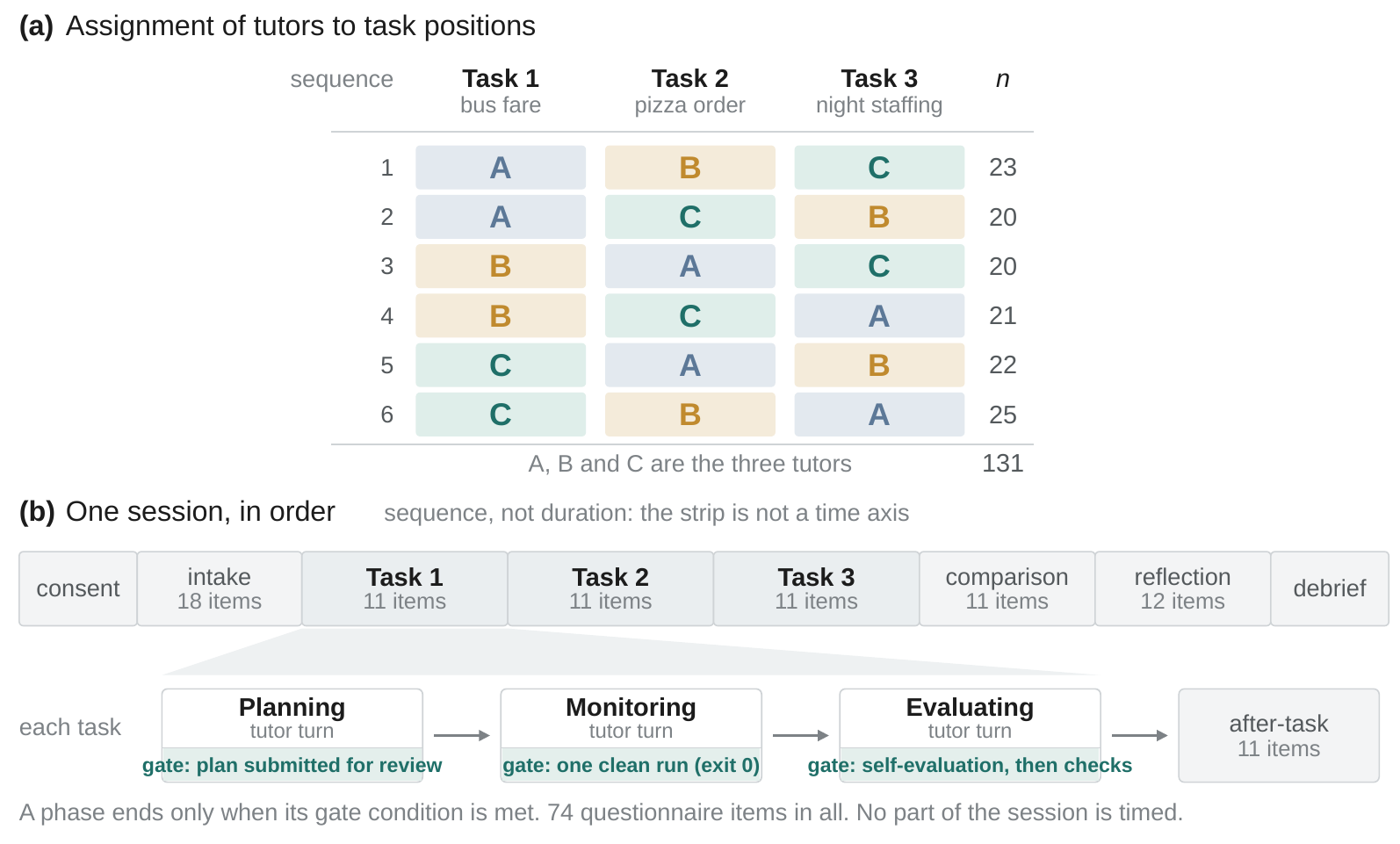}
\caption{Design and procedure. (a) The six tutor orders, one per sequence, with the
realised allocation \emph{n}. (b) One session in the order met, with
questionnaire item counts, and the three gated phases of a task. The strip
is a sequence, not a time axis; no part of the session was timed.}
\Description{Two panels. (a) A six-row table. Each row is one of the six
orders in which the three tutors can be met, shown as tinted cells lettered A,
B or C under the headings Task 1, Task 2 and Task 3, with the number of
participants assigned to that order at the right, summing to 131. (b) A strip of
eight blocks in order: consent, intake, the three tasks, comparison,
reflection and debrief, each block carrying its questionnaire item count.
A wedge widens from the Task 1 block into an expansion below, showing the
three phases of one task --- Planning, Monitoring and Evaluating --- each box
carrying at its foot the gate condition that ends that phase: plan submitted
for review, one clean run, self-evaluation then checks. The last arrow leads
to the after-task questionnaire.}
\label{fig:timeline}
\end{figure*}

Sessions followed a fixed sequence (Figure~\ref{fig:timeline}b): briefing and
informed consent, login, an 18-item intake, the three tasks, each run in three
gated phases and each closed by an eleven-item after-task questionnaire, then
the comparative set, an exit reflection, and debriefing. No part was timed, and
participants could spend as long as they wished on any task. Every instrument
reading is taken from its first sitting, and repeats produced by a resume are
dropped. Completion is defined by a passing final submission rather than by the
session status field. Participants who completed the session received
US\$10.

\subsection{Analysis for RQ1 to RQ3}

Estimates are differences from \armC{}, positive where \armC{} is higher,
with two planned contrasts, \armC{} against \armA{} and against \armB{}, and
task and position as covariates throughout. Holm correction is applied within
each family, the contrasts serving one question, five after-task items under
RQ1, five under RQ2, frustration under RQ3, and corrected values are reported
as $p_{\mathrm{Holm}}$; it holds the chance of a false positive across a
family at 5\% without assuming the tests are independent, which they are
not, since every item comes from the same participants. After-task items
are modelled with linear mixed models with a random intercept for
participant, since each participant supplies one reading per tutor; where a
hypothesis predicts no difference, a non-significant contrast is not treated
as support, and equivalence is tested with two one-sided tests (TOST,
reported as $p_{\mathrm{TOST}}$) against a bound of 0.25 scale points, the
effect the sample is powered to detect.
Session-level binary outcomes and counts use generalised estimating equations
clustered on participant with logit and log links, which give
population-averaged rates and ratios with standard errors that respect the
three sessions per participant. Comparative items are
tested against equal attribution with Holm correction over the ten items
administered. Analyses used Python 3.11,
pandas 3.0.2, statsmodels 0.15.0 and scipy 1.17.1.

\subsection{Annotation for RQ4 on fading episodes}

We content-analyse what learners supplied at each fading, then test those codes
against an outcome the coding cannot influence. Tutor turns are model
completions of authored rung instructions, so coding is confined to learner
turns. The unit is the fading episode: one moment at which \armC{} judged the learner
to have taken support up and returned the ladder to its lightest rung. \armC{}
faded on 551 occasions across 106 sessions; coding covers 515 episodes from 100
sessions, the six excluded sessions dropped whole rather than sampled away.
Each episode was reconstructed from the log as a fixed record: rung before the
drop, phase, open decision points, the tutor's support turn immediately before,
every learner turn in the window, and any workspace event in it.
\begin{table*}[t]
\footnotesize
\centering
\caption{Warrant codes for fading episodes, ordered from highest to lowest
warrant, each with one learner turn from the corpus. Each episode receives
exactly one code, the highest that applies.}
\label{tab:warrant}
\setlength{\tabcolsep}{6pt}
\begin{tabular}{@{}>{\raggedright\arraybackslash}p{0.15\textwidth}>{\raggedright\arraybackslash}p{0.34\textwidth}>{\raggedright\arraybackslash}p{0.45\textwidth}@{}}
\toprule
Code & What the learner supplied & Example \\
\midrule
\multicolumn{3}{@{}l}{\textbf{Take-up evidenced}}\\
Demonstration & The substance of a decision, correct or not &
  \emph{Base fare first, 0.60 for children and seniors, 1.40 for adults; then the peak surcharge, adults only.} \\
Artifact & A pasted plan, program or run result, with no articulation &
  \texttt{age = input("Enter age: ")} \texttt{hour = input("Enter hour (0--23): ")} \\
\addlinespace[2pt]
\multicolumn{3}{@{}l}{\textbf{Take-up not evidenced}}\\
Assertion & A claim to be done or to understand, with no task content &
  \emph{All three test cases have run and the output matches.} \\
Request or resistance & Asks the tutor to decide, or pushes back; reasoned argument against the tutor codes as demonstration &
  \emph{Can I start writing the code now? What else needs changing in my plan?} \\
Wrong decision & Substantive content, but not about what was asked or open &
  Asked how to check the card input: \emph{if card == y: price -= 0.20. Is this plan OK?} \\
\addlinespace[2pt]
\multicolumn{3}{@{}l}{\textbf{Outside the aim contrast}}\\
Procedural & About the interface &
  \emph{Where is the editor? I only have the terminal and the plan box.} \\
None & No learner turn &
  The window held a code run and no message. \\
\bottomrule
\end{tabular}
\end{table*}

The scheme was developed inductively from a calibration read of 24 episodes;
each episode receives the highest-warrant code that applies
(Table~\ref{tab:warrant}), and Appendix~\ref{app:coding} gives the full
codebook and what the second coder received.
A second dimension is theory-driven and is the one RQ4 turns on: contingency
requires a diagnosis of the learner's state~\cite{vandepol2010scaffolding}, and
the open question is what evidence that diagnosis rests on. Each episode
therefore also carries \emph{aim}: true where the learner's content addresses
something open or asked, false otherwise, undefined for procedural episodes and
those with no learner turn, which puts 28 episodes outside the contrast. By
rule, a request that the tutor settle the decision under support counts as
aimed, since it names that decision. Lastly, regarding the annotation process, we have one annotator coded all 515 episodes, and a second annotator,
blind to those codes and to session order, coded a stratified sample of 103 of
the 509 episodes assigned by judgement (20\%). Agreement was substantial:
92.2\% on warrant, Cohen's $\kappa = .74$, 95\% CI $[.68, .80]$; 94.2\% on aim,
$\kappa = .69$, $[.55, .81]$. 

\section{Quantitative Findings on Tutor Comparison}
\label{sec:results}

This section answers RQ1 to RQ3 by comparing the three tutors on two
records: what learners reported after each task (Table~\ref{tab:aftertask})
and what the event log shows they and the tutor did
(Table~\ref{tab:process}). RQ1 reads demand and perceived load from the
self-report items and the forced-choice attributions; RQ2 reads delegation
and delivery from the log and ownership from self-report; RQ3 reads
frustration from both. Every estimate is a contrast against \armC{},
positive where \armC{} is higher, fitted and corrected as
Section~\ref{sec:method} states.

\subsection{RQ1: demand rises, load does not}

Table~\ref{tab:aftertask} gives the after-task means and the two planned
contrasts. \armC{} exceeded \armA{} on effort ($+0.46$, $SE$ 0.09,
$p_{\mathrm{Holm}} < .001$) and on effort spent on the approach rather than
the typing ($+0.33$, $SE$ 0.09, $p_{\mathrm{Holm}} < .001$). Reports of
having worked out the approach before coding run the same way ($+0.24$,
$SE$ 0.09) but do not survive correction ($p_{\mathrm{Holm}} = .051$), so we
record the direction and claim nothing from it. Against \armB{}, none of the three
differ, and for two of them the absence of a difference is demonstrable
rather than merely unrejected: approach effort is equivalent within $0.25$
scale points (C\,--\,B $= +0.03$, 95\% CI $[-0.14, +0.19]$,
$p_{\mathrm{TOST}} = .004$), as is planning first ($+0.01$, $[-0.16, +0.18]$,
$p_{\mathrm{TOST}} = .003$). Effort itself is not: \armC{} sits $0.11$ above
\armB{}, and the interval $[-0.06, +0.29]$ still admits a difference at the
bound ($p_{\mathrm{TOST}} = .059$).

\begin{table*}[t]
\footnotesize
\centering
\caption{After-task self-report ($N = 131$ participants, 393 sessions). Means by tutor and mixed-effects contrasts against
\armC{} (positive = \armC{} higher), with Holm-corrected $p$ within each
family. All items 1--5 except tutor authorship (1--4, higher = more of the
work done by the tutor). Bold contrasts survive correction; the asterisks carry the same
information.}
\label{tab:aftertask}
\setlength{\tabcolsep}{8pt}
\resizebox{0.75\textwidth}{!}{%
\begin{tabular}{@{}lccc rr@{}}
\toprule
& \armC{} & \armA{} & \armB{} & \multicolumn{1}{c}{C\,--\,A ($SE$)} & \multicolumn{1}{c}{C\,--\,B ($SE$)} \\
\midrule
\multicolumn{6}{@{}l}{\textbf{RQ1 \textperiodcentered\ Metacognitive demand and cognitive load}}\\
Effort                & 3.44 & 2.98 & 3.34 & $\mathbf{+0.46}$*** {\scriptsize(0.09)} & $+0.11$ {\scriptsize(0.09)} \\
Effort on approach    & 3.28 & 2.95 & 3.27 & $\mathbf{+0.33}$*** {\scriptsize(0.09)} & $+0.03$ {\scriptsize(0.09)} \\
Planned first         & 3.49 & 3.25 & 3.48 & $+0.24$ {\scriptsize(0.09)} & $+0.01$ {\scriptsize(0.09)} \\
Perceived task load   & 3.36 & 3.30 & 3.29 & $+0.07$ {\scriptsize(0.08)} & $+0.08$ {\scriptsize(0.08)} \\
Germane load          & 3.85 & 3.74 & 3.78 & $+0.11$ {\scriptsize(0.08)} & $+0.07$ {\scriptsize(0.08)} \\
\addlinespace[2pt]
\multicolumn{6}{@{}l}{\textbf{RQ2 \textperiodcentered\ Delegation and delivery}}\\
Direct help received  & 3.21 & 3.53 & 3.00 & $\mathbf{-0.33}$** {\scriptsize(0.10)} & $+0.20$ {\scriptsize(0.10)} \\
Help gap (wanted vs got) & 3.63 & 3.44 & 3.44 & $+0.20$ {\scriptsize(0.08)} & $+0.19$ {\scriptsize(0.08)} \\
Tutor authorship      & 2.30 & 2.63 & 2.40 & $\mathbf{-0.33}$*** {\scriptsize(0.08)} & $-0.10$ {\scriptsize(0.08)} \\
Could do it alone     & 3.33 & 3.13 & 3.15 & $+0.20$ {\scriptsize(0.08)} & $+0.18$ {\scriptsize(0.08)} \\
Understands the solution & 3.66 & 3.47 & 3.40 & $+0.20$ {\scriptsize(0.08)} & $\mathbf{+0.26}$* {\scriptsize(0.08)} \\
\addlinespace[2pt]
\multicolumn{6}{@{}l}{\textbf{RQ3 \textperiodcentered\ Frustration}}\\
Frustration           & 2.44 & 2.02 & 2.73 & $\mathbf{+0.42}$*** {\scriptsize(0.11)} & $\mathbf{-0.29}$* {\scriptsize(0.11)} \\
\bottomrule
\multicolumn{6}{@{}l}{\scriptsize *$p_{\mathrm{Holm}}<.05$\quad
**$p_{\mathrm{Holm}}<.01$\quad ***$p_{\mathrm{Holm}}<.001$}
\end{tabular}%
}
\end{table*}
Furthermore, neither perceived-load item separates any pair of tutors, and both are
equivalent between \armC{} and \armB{} within the same bound: perceived task
load sits at 3.36, 3.30 and 3.29 ($p_{\mathrm{TOST}} = .020$) and germane
load at 3.85, 3.74 and 3.78 ($p_{\mathrm{TOST}} = .007$). Learners also named \armC{} most often as the
tutor with which they thought hardest (52 of 131, $\chi^2(2) = 16.8$,
$p_{\mathrm{Holm}} = .002$) and did the most thinking themselves (49,
$\chi^2(2) = 13.2$, $p_{\mathrm{Holm}} = .011$), with \armB{} second and
\armA{} a distant third on both; the full set of ten forced-choice items
is in Appendix~\ref{app:comparative}. Therefore, demand moved toward the level of the tutor that withholds
structuring while load stayed where it was, the pattern H1 predicted, with
two qualifications: forethought shows the direction without significance,
and effort is the one demand item not shown equivalent to \armB{}. \noindent\textbf{In short, tutor C placed the same metacognitive demand as Tutor B and more than Tutor A; perceived load did not differ between tutors.}

\begin{table*}[t]
\footnotesize
\centering
\caption{Behaviour from the event log, per session ($N = 131$, 393 sessions).
Rates are the share of sessions in which the event occurred; turns, lines,
programs and runs are means. Contrasts are population-averaged GEE estimates clustered on
participant, with task and position as covariates: odds ratios ($OR$) for
the rates, rate ratios ($RR$) for the counts, each stated as \armC{} against
the named tutor.}
\label{tab:process}
\setlength{\tabcolsep}{6pt}
\resizebox{\textwidth}{!}{%
\begin{tabular}{@{}lccc ll@{}}
\toprule
& \armC{} & \armA{} & \armB{} & \multicolumn{1}{c}{C vs A} & \multicolumn{1}{c}{C vs B} \\
\midrule
\multicolumn{6}{@{}l}{\textbf{Delegation}}\\
Asked the tutor to build & 50.4\% & 51.1\% & 23.7\% &
$OR = 0.95$ $[0.64, 1.42]$, $p = .82$ & $OR = 3.29$ $[1.98, 5.48]$, $p < .001$ \\
Learner turns            & 16.8 & 5.0 & 14.0 &
$RR = 3.30$ $[2.72, 3.99]$, $p < .001$ & $RR = 1.16$ $[0.94, 1.44]$, $p = .17$ \\
\addlinespace[2pt]
\multicolumn{6}{@{}l}{\textbf{Delivery}}\\
Tutor-written artifact delivered & 48.1\% & 23.7\% & \phantom{0}9.2\% &
$OR = 2.99$ $[1.80, 4.98]$, $p < .001$ & $OR = 9.71$ $[5.29, 17.85]$, $p < .001$ \\
Lines of code the tutor wrote & 25.3 & 15.4 & \phantom{0}4.5 &
$RR = 1.64$ $[1.26, 2.13]$, $p < .001$ & $RR = 5.57$ $[3.93, 7.90]$, $p < .001$ \\
Complete programs written in chat & 0.89 & 0.84 & 0.19 &
$RR = 1.06$ $[0.71, 1.57]$, $p = .77$ & $RR = 4.60$ $[2.79, 7.58]$, $p < .001$ \\
\addlinespace[2pt]
\multicolumn{6}{@{}l}{\textbf{Execution}}\\
Runs logged as failing   & 4.06 & 2.37 & 5.44 &
$RR = 1.68$ $[1.26, 2.25]$, $p < .001$ & $RR = 0.71$ $[0.56, 0.91]$, $p = .007$ \\
Runs exiting with an error & 2.46 & 1.08 & 3.45 &
$RR = 2.18$ $[1.36, 3.49]$, $p = .001$ & $RR = 0.67$ $[0.45, 0.99]$, $p = .047$ \\
\bottomrule
\end{tabular}%
}
\end{table*}
\subsection{RQ2: delegation holds, delivery rises}

Learners asked \armC{} to build as often as they asked \armA{}
(Table~\ref{tab:process}). At least one build request occurred in 50.4\% of
\armC{} sessions and 51.1\% of \armA{} sessions ($OR = 0.95$,
95\% CI $[0.64, 1.42]$, $p = .82$), against 23.7\% of \armB{} sessions
($OR = 3.29$ $[1.98, 5.48]$, $p < .001$). The demand \armC{} places did not
deter delegation; the structuring \armB{} withholds did, halving it. The
interval on the \armC{} against \armA{} odds ratio excludes any reduction
larger than 36\%, so the claim is that contingency does not substantially
deter delegation rather than that it leaves it untouched.

What the tutors then did with those requests separates them, and not in the
direction the design vocabulary predicts. A tutor-written artifact was
delivered into the workspace in 48.1\% of \armC{} sessions, against 23.7\%
for \armA{} ($OR = 2.99$ $[1.80, 4.98]$, $p < .001$) and 9.2\% for \armB{}
($OR = 9.71$ $[5.29, 17.85]$, $p < .001$). The authorship probe agrees with the log flag ($\kappa = .76$). The amount points the same way: the tutor wrote
more code per session under \armC{} than under \armA{} ($RR = 1.64$
$[1.26, 2.13]$, $p < .001$), and it came in pieces, since complete programs
written into the chat were as frequent under \armA{} as under \armC{}
($RR = 1.06$, $p = .77$). The tutor that withholds until a decision is settled
carried more of the labour than the tutor that never withholds at all, and
carried it one decision at a time. And the exchange around that transfer differs sharply. Learners produced a mean
of 16.8 turns per session with \armC{} and 14.0 with \armB{}, statistically
indistinguishable ($RR = 1.16$ $[0.94, 1.44]$, $p = .17$), against 5.0 with
\armA{} ($RR = 3.30$ $[2.72, 3.99]$, $p < .001$). Runs logged as
failing, meaning a run that raised an error or did not pass its tests,
followed the same shape but not the same order: 4.06 per session under
\armC{}, 2.37 under \armA{} and 5.44 under \armB{}.

On self-report, learners reported receiving less direct help from \armC{} than from
\armA{} ($-0.33$, $p_{\mathrm{Holm}} = .009$) and attributed the finished
solution less to the tutor ($-0.33$, $p_{\mathrm{Holm}} < .001$). Two
related items, the help gap and being able to do it alone, move in \armC{}'s
favour without surviving correction ($p_{\mathrm{Holm}} = .097$), and on
understanding of the submitted solution \armC{} sits above \armB{} by $0.26$
($p_{\mathrm{Holm}} = .014$). H2 is supported:
contingency changed what a learner had to settle before help arrived, not
whether it arrived. \noindent\textbf{In short, elegation to Tutor C was as frequent as to Tutor A; delivery by Tutor C was higher than by Tutor A, and came one decision at a time rather than as whole programs.}

\subsection{RQ3: frustration sits between the two baselines}

Frustration was higher under \armC{} than under \armA{} ($+0.42$,
$p_{\mathrm{Holm}} < .001$) and lower than under \armB{} ($-0.29$,
$p_{\mathrm{Holm}} = .011$). \armC{} therefore sits between the two
baselines, above the tutor that answers on request and below the tutor that
withholds structuring, which is what H3 predicted in both directions. The forced-choice item is weaker on the same point: asked which tutor was
most frustrating, learners named \armC{} and \armB{} equally often (40 each,
against 18 for \armA{}, $p_{\mathrm{Holm}} = .050$). The two instruments
agree that \armA{} is the least frustrating and differ on whether \armC{} and
\armB{} can be separated. \noindent\textbf{In short, frustration under Tutor C was higher than under Tutor A but lower than under Tutor B.}

\section{Qualitative Findings on Fading Warrant}
\label{sec:warrant}

The contingent shift rule fades support on success, and in the tutoring it was written for, success is the learner answering the tutor's question (Section 2.3). An AI tutor cannot count on that answer. After Tutor C's first ask, the learner's next turn resolved it in 37.0\% of rounds, and the rate fell with each further ask, to 21.8\% after the third (Figure 7b). Fading still came early: of the rounds learners took up, nine in ten were within the three asks and three in five at the first, so naming the decision was usually the whole intervention. But it was not usually answered. Of the 515 coded fading episodes, only 282 followed an answer to the question. The rest followed something else, most often a pasted plan or program: the learner had fixed the thing rather than replied. The remainder were a bare it's fine, a request that the tutor build, a remark about the interface, or content about a different decision. Where the learner did answer, the answer often did not settle the decision: the tutor later conceded 30.3\% of the decisions still open after an answer. Nearly half of the tutor's fading rested on a turn that was not an answer, and a rule that fades on a correct answer has nothing to say about them, nor about the answers that left the decision open. Hence, the rest of this section shows what did warrant fading, and then what a tutor that fades on that warrant no longer has to do.

\begin{figure*}[t]
\centering
\includegraphics[width=\textwidth]{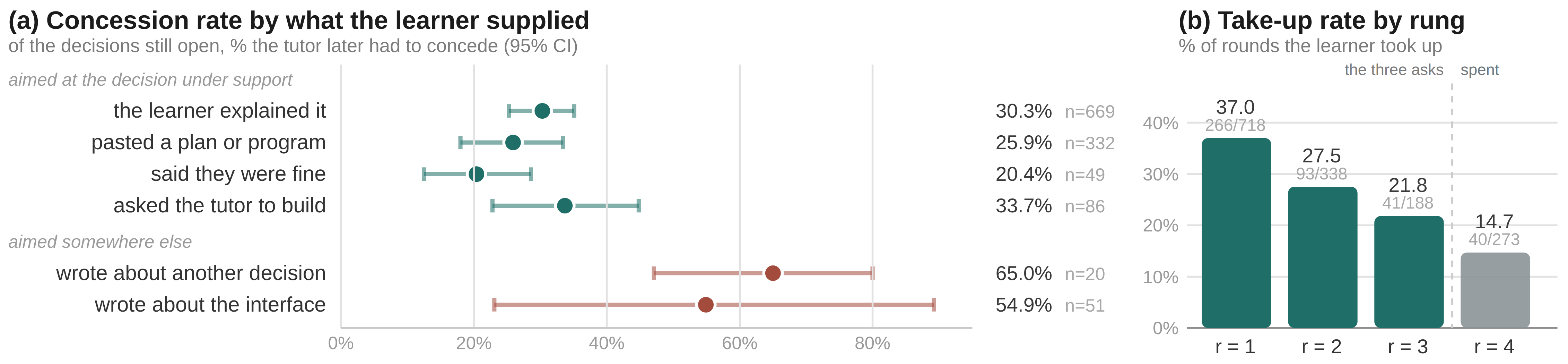}
\caption{(a) Share of decisions still open at fading that the tutor later
conceded, by what the learner had supplied; 95\% bootstrap intervals
resampled on participant, rows grouped by aim. (b) Take-up rate by rung;
the first three bars are the tutor's asks, the last the spent ladder, where
the point has already been conceded.}
\Description{(a) A dot-and-interval chart of concession rate by warrant code,
in two groups. Aimed at the decision under support: the learner explained it
30.3 percent (95 percent interval 25.3 to 35.1, 669 open decisions), pasted a
plan or program 25.9 (18.0 to 33.4, 332), said they were fine 20.4 (12.5 to
28.6, 49), asked the tutor to build 33.7 (22.8 to 44.8, 86). Aimed somewhere
else: wrote about another decision 65.0 (47.1 to 80.0, 20), wrote about the
interface 54.9 (23.1 to 89.2, 51). The first four intervals overlap one
another; the interval for another decision lies above them, while the
interface interval is too wide to separate. (b) A bar chart of take-up rate by rung,
with a dashed line after the third bar: the tutor's three asks at 37.0
percent of 718 rounds, 27.5 percent of 338 and 21.8 percent of 188, then
rounds on a spent ladder, where the tutor concedes and builds, at 14.7
percent of 273.}
\label{fig:warrant}
\end{figure*}

\subsection{Aim is the warrant}

Whether fading held is in the log: a point faded from and later conceded is
one the tutor got wrong. Scored that way (Figure~\ref{fig:warrant}a), the
turns that differ in depth do not order by depth. After a full demonstration the tutor later conceded 30.3\% of what was
still open, after a pasted artifact 25.9\%, after a bare assertion 20.4\%,
and after a request to build 33.7\%: the shallowest lowest. A learner who fixed the plan without replying was as safe to fade
from as one who explained the rule in full, and one who said \emph{it's
fine} safer than either. What separates is aim. Content about a
\emph{different} decision was followed by concession 65.0\% of the time,
twice the rate of anything aimed at the decision under support, and the gap holds within
rungs, so it is not an artifact of where on the ladder the turn fell. The
aim dimension as a whole says the same: turns not aimed at the decision
under support were followed by concession 40.3\% of the time (72 open
decisions) against 28.8\% for aimed turns (1,084), a difference of 11.5
points, 95\% bootstrap interval on participant $[2.1, 22.4]$.

The reason is in what the ask already did. The ask itself gives
information: it names the decision and points to where the answer lives. So
the learner's reply mainly shows whether the ask reached them, not how much
they understood: a fixed plan or a short confirmation about the right
decision tells the tutor its ask landed; an elaborate answer about a
different decision tells it the ask did not.

\begin{quote}\sloppy
\textbf{Fading Warrant for AI Tutor.} Contingent fading requires the learner's turn to be
aimed at the decision under support. How much the turn says does not decide
whether fading holds. Whether it is about the right decision does.
\end{quote}

Contingency therefore has three parts: direction, the contingent shift
rule of Section~\ref{sec:related:contingency}; unit, the scoping of support
to decision points in Section~\ref{sec:theory}; and warrant, what the
take-up test of Equation~\ref{eq:rung} left open.

\subsection{Contingent scaffolding surrenders less}

\begin{table*}[t]
\footnotesize
\centering
\caption{What it took to get the full answer ($N = 131$ participants, 393 sessions).
Each cell is the share of sessions in which the event occurred at any point;
\emph{full answer surrendered} is an end state, the whole answer given away,
not the delivery rate of Table~\ref{tab:process}, which counts a session as
soon as the tutor places any artifact, whole or partial. A build request is the
same act under every tutor, but \armA{} builds what is asked and no more, so
its cell counts sessions in which it produced the complete program, in chat or
in the workspace; \armB{} and \armC{} surrender only through the welfare floor.
$p$: McNemar exact, paired within participant, against \armC{}.}
\label{tab:floor}
\setlength{\tabcolsep}{9pt}
\begin{tabular}{@{}l rr rr rr@{}}
\toprule
& \multicolumn{2}{c}{Asked the tutor to build} & \multicolumn{2}{c}{Full answer surrendered} & \multicolumn{2}{c}{Distress detected} \\
\cmidrule(lr){2-3}\cmidrule(lr){4-5}\cmidrule(lr){6-7}
Tutor & \% & $p$ & \% & $p$ & \% & $p$ \\
\midrule
\armA{} & 51.1 & $1.00$ & \textbf{59.5} & $\mathbf{<.001}$ & \phantom{0}\textbf{0.8} & $\mathbf{.006}$ \\
\armB{} & \textbf{23.7} & $\mathbf{<.001}$ & \textbf{16.0} & $\mathbf{.007}$ & \textbf{16.0} & $\mathbf{.021}$ \\
\armC{} & 50.4 & \textendash{} & \phantom{0}6.1 & \textendash{} & \phantom{0}8.4 & \textendash{} \\
\bottomrule
\end{tabular}
\end{table*}


After understanding how tutor C approach fading, we asked what let \armC{} withhold as much as \armB{} does without
driving learners to give up. Every tutor surrenders the full answer in some sessions; what differs is what
it takes (Table~\ref{tab:floor}). The measure is a session's end state,
whether the whole answer was given away, and so is stricter than the
delivery of Section~\ref{sec:results}, which counts a session as soon as
the tutor places any artifact, a partial one included. Learners asked \armC{} to
build as often as they asked \armA{}, so the demand was the same; what
differed is what the request got them. \armA{} produced the complete
program in three sessions in five. \armB{} refused, and released only when the learner gave up, which one
session in six reached. \armC{} put the request on the ladder, asking about
the decisions the build turned on, and the asks ran out in one session in
sixteen; the floor fired under \armB{} more than twice as often as under
\armC{}. Distress
followed the same order, twice as common under \armB{} as under \armC{} and
almost absent under \armA{}.

What separates \armC{} from \armB{} is therefore not how much it withholds
but what it does when withholding stops working. \armB{} has one move, the
release with no return that the Socratic gate also
makes~\cite{sun2026socratic}. \armC{} concedes one point, keeps the rest of
the coverage set, and returns to $r = 1$. The same numbers carry a second
reading. A floor that opens on a statement of giving up is also a route for
gaming the system, the help abuse in which a learner performs the trigger to
obtain what is withheld~\cite{baker2004offtask,aleven2016help}, and the log
cannot separate a strategic surrender from a sincere one. On either reading
\armC{} is the tutor under which that route was taken least, because the ladder
gives a learner who wants the answer something cheaper to do than declare
defeat: answer the ask. 

\textbf{In short, fading is warranted when the learner's turn is aimed at the decision
under support, whatever its depth; a turn about another decision is not a
warrant. Fading on this warrant, \armC{} surrendered the full answer in 1
session in 16, against 1 in 6 under \armB{}.}

\section{Discussion}
\label{sec:discussion}

\subsection{Demand and labour come apart}

If the only thing that mattered were how much a tutor gives away, giving
less would always mean demanding more, and every design would sit on one
line. The three tutors do not. \armC{} placed the demand of the tutor that withholds structuring
and carried more of the labour than the tutor that supplies it on request:
effort on the approach is equivalent to \armB{}'s within a quarter of a
scale point, while an artifact reached the workspace in 48.1\% of \armC{}
sessions against 23.7\% under \armA{}, and neither perceived-load item
separates any pair. The claim is about what learners reported and wrote,
not about intrinsic or extraneous load, which self-report does not separate. On one axis that looks like a contradiction. On two axes it is exactly the
position the principle describes, and the paper's first claim is that a
tutor can sit there: what \armC{} varies is not how much help arrives but
what has to be true before it does~\cite{kapur2016examining}. Restricting an assistant is not the
only way to keep the analytic work with the learner, and on this evidence not
the most effective one, since the tutor that restricted most is also the one
learners delegated to least.

\subsection{The price of demand}

Demand is not free. \armC{} was 0.42 more frustrating than \armA{}, and
productive failure predicts exactly that~\cite{kapur2008productive}. The finding worth attention is the other
direction: \armC{} was 0.29 \emph{less} frustrating than \armB{}, so
contingency recovers part of the cost that withholding incurs without giving
up the demand. The mechanism is in the design: when support fades on take-up,
control returns to a learner who has shown they can carry the decision. That
is the resolution which confusion research treats as the difference between
productive and unproductive difficulty~\cite{dmello2012dynamics}. On a single forced choice learners named \armC{} and
\armB{} equally often as most frustrating, where the after-task rating
separates them; a design that raises demand will have an emotional cost that can be
measured, and this one does.

\subsection{What a fading system must read}

Contingency is said to require reading the learner's
state~\cite{vandepol2010scaffolding}. What exactly is read was never
spelled out, because a human tutor beside the learner can see the attempt
succeed or fail~\cite{wood1976role}. A tutor on a text channel must answer a
question the literature never had to pose, and the answer here is that the
learner's turn must be aimed at the decision under support, and how much
content it carries does not matter (Section~\ref{sec:warrant}). The
consequence is a specification: \armC{} decides take-up
by pattern, judgement and length, all measures of depth, and the evidence
says the measure should be aim. That is checkable in the next version, since the
two come apart exactly where fading fails, and it applies to any
system that fades.

\subsection{Design implications}

Four implications follow. \emph{Make release contingent; do not cap
assistance.} Restricting what a tutor may say sets one position for every
learner; making release contingent on a settled decision sets the
demand, lets the labour vary, and does not cost delegation. \emph{Measure
aim, not effort.} A take-up detector should ask whether the turn addresses
the decision under support; a careful answer about the wrong decision is the
case a depth measure most confidently gets wrong. \emph{Reset on success.}
Fading is what separates working to someone else's standard from
regulating one's own~\cite{schunk1997social}, and it is where the emotional
saving comes from. \emph{Instrument the record, and give a withholding design
one door.} Counting what a tutor is willing to say misses what reaches the
workspace and in what pieces: complete programs in the chat were as frequent
under \armA{} as under \armC{}, while lines written and artifacts placed
were not. A design that refuses needs a floor that opens on an
explicit statement of giving up rather than on frustration, or the
manipulation dissolves for the learners who most need the demand; ours fired
28 times in 21 of \armB{}'s 131 sessions.

\section{Scope and Further Work}
\label{sec:limitations}

The study fixes several things deliberately in order to isolate contingency,
and each choice marks out the study that follows.

\textbf{One sitting, completion held constant.} Participants met all three
tutors in one sitting, untimed, and the analysed sample is by construction
those who passed every task, which is the right unit for process measures
and what makes the tutors comparable on demand and delivery rather than on
whether learners finished.
It is short against the timescale on which knowledge consolidates, so whether
contingent support affects retention is the question this design is least
able to answer; the study that answers it runs \sys{} across a course with a
delayed post-test taken without the tool and a transfer task that lets
outcomes vary.

\textbf{Load measurement.} Load is measured by self-report and effort by a
single item, so intrinsic and extraneous load are not separated, and germane
load is corroborated by the learner's written responses rather than by a
second instrument. A study with dual-task or physiological measures could
test whether contingency also changes the load partition, which this design
cannot see.

\textbf{Population.} Participants were adult learners who answered a social
media call with generative AI already
routine in their coursework. Whether the design holds
where AI use is less established is a question about populations rather than
about the mechanism, and it is answered by running the same three tutors
elsewhere.

\section{Conclusion}
\label{sec:conclusion}

Discussion of AI in learning treats the demand a tutor places on a learner
as whatever is left once assistance is withheld. Separating that demand from
the cognitive load a system carries gives a second axis, the contingency
dimension; we took \emph{preserved metacognitive demand} as the principle it
implies and built \sys{} to occupy three points on it. Across 131 adult learners and three
Python tasks, the contingent tutor matched the tutor that withholds
structuring on the effort learners spent on the approach, with no difference
in perceived load, while delivering an artifact twice as often as the tutor that
answers on request and leaving delegation undisturbed; it was more
frustrating than the tutor that answers and less than the tutor that
withholds. Its fading held when the learner's turn was aimed
at the decision under support, not when it was elaborate, which names the
property a fading system must read; fading on that warrant left the full
answer surrendered in one session in sixteen, against one in six under the
tutor that withholds. Restricting an assistant is not the only
way to keep the analytic work with the learner, and on this evidence not the
most effective one: a tutor can be free in what it says, as long as what the
learner must settle before it speaks is held fixed.

\begin{acks}
\textbf{Generative AI use.} The three tutors share a large language model backbone; Section~\ref{sec:system}
states what each may say. In preparing this paper, the authors used Claude
Fable 5.1 for coherence and consistency checks, for critical review of drafts,
and for grammar correction. All analyses, claims, figures and final wording
were specified, verified and edited by the authors, who take responsibility for
them.
\end{acks}

\bibliographystyle{ACM-Reference-Format}
\bibliography{refs}

@inproceedings{finnieansley2022robots,
  author    = {Finnie-Ansley, James and Denny, Paul and Becker, Brett A. and
               Luxton-Reilly, Andrew and Prather, James},
  title     = {The Robots Are Coming: Exploring the Implications of {OpenAI}
               {Codex} on Introductory Programming},
  booktitle = {Proceedings of the 24th Australasian Computing Education
               Conference (ACE '22)},
  year      = {2022}, pages = {10--19}, publisher = {ACM},
  doi       = {10.1145/3511861.3511863},
  address = {New York, NY, USA}
}

@inproceedings{finnieansley2023myai,
  author    = {Finnie-Ansley, James and Denny, Paul and Luxton-Reilly, Andrew and
               Santos, Eddie Antonio and Prather, James and Becker, Brett A.},
  title     = {My {AI} Wants to Know if This Will Be on the Exam: Testing
               {OpenAI}'s {Codex} on {CS2} Programming Exercises},
  booktitle = {Proceedings of the 25th Australasian Computing Education
               Conference (ACE '23)},
  year      = {2023}, pages = {97--104}, publisher = {ACM},
  doi       = {10.1145/3576123.3576134},
  address = {New York, NY, USA}
}

@inproceedings{savelka2023thrilled,
  author    = {Savelka, Jaromir and Agarwal, Arav and An, Marshall and
               Bogart, Chris and Sakr, Majd},
  title     = {Thrilled by Your Progress! Large Language Models ({GPT-4}) No
               Longer Struggle to Pass Assessments in Higher Education
               Programming Courses},
  booktitle = {Proceedings of the 2023 ACM Conference on International Computing
               Education Research (ICER '23), Volume 1},
  year      = {2023}, pages = {78--92}, publisher = {ACM},
  doi       = {10.1145/3568813.3600142},
  address = {New York, NY, USA}
}

@article{daniotti2026who,
  author  = {Daniotti, Simone and Wachs, Johannes and Feng, Xiangnan and
             Neffke, Frank},
  title   = {Who is using {AI} to code? Global diffusion and impact of
             generative {AI}},
  journal = {Science}, volume = {391}, number = {6787}, pages = {831--835},
  year    = {2026}, doi = {10.1126/science.adz9311}
}

@article{bird2023taking,
  author  = {Bird, Christian and Ford, Denae and Zimmermann, Thomas and
             Forsgren, Nicole and Kalliamvakou, Eirini and Lowdermilk, Travis and
             Gazit, Idan},
  title   = {Taking Flight with {Copilot}},
  journal = {Communications of the ACM}, volume = {66}, number = {6},
  pages   = {56--62}, year = {2023}, doi = {10.1145/3589996}
}

@inproceedings{becker2023programming,
  author    = {Becker, Brett A. and Denny, Paul and Finnie-Ansley, James and
               Luxton-Reilly, Andrew and Prather, James and Santos, Eddie Antonio},
  title     = {Programming Is Hard -- Or at Least It Used to Be: Educational
               Opportunities and Challenges of {AI} Code Generation},
  booktitle = {Proceedings of the 54th ACM Technical Symposium on Computer
               Science Education (SIGCSE TS '23), Volume 1},
  year      = {2023}, pages = {500--506}, publisher = {ACM},
  doi       = {10.1145/3545945.3569759},
  address = {New York, NY, USA}
}

@article{denny2024computing,
  author  = {Denny, Paul and Prather, James and Becker, Brett A. and
             Finnie-Ansley, James and Hellas, Arto and Leinonen, Juho and
             Luxton-Reilly, Andrew and Reeves, Brent N. and
             Santos, Eddie Antonio and Sarsa, Sami},
  title   = {Computing Education in the Era of Generative {AI}},
  journal = {Communications of the ACM}, volume = {67}, number = {2},
  pages   = {56--67}, year = {2024}, doi = {10.1145/3624720}
}

@inproceedings{vadaparty2024cs1llm,
  author    = {Vadaparty, Annapurna and Zingaro, Daniel and Smith IV, David H. and
               Padala, Mounika and Alvarado, Christine and
               Benario, Jamie Gorson and Porter, Leo},
  title     = {{CS1-LLM}: Integrating {LLMs} into {CS1} Instruction},
  booktitle = {Proceedings of the 2024 on Innovation and Technology in Computer
               Science Education (ITiCSE '24), Volume 1},
  year      = {2024}, pages = {297--303}, publisher = {ACM},
  doi       = {10.1145/3649217.3653584},
  address = {New York, NY, USA}
}

@article{wing2006computational,
  author  = {Wing, Jeannette M.},
  title   = {Computational thinking},
  journal = {Communications of the ACM}, volume = {49}, number = {3},
  pages   = {33--35}, year = {2006}, doi = {10.1145/1118178.1118215}
}

@article{grover2013computational,
  author  = {Grover, Shuchi and Pea, Roy},
  title   = {Computational Thinking in {K--12}: A Review of the State of the Field},
  journal = {Educational Researcher}, volume = {42}, number = {1},
  pages   = {38--43}, year = {2013}, doi = {10.3102/0013189X12463051}
}

@article{shute2017demystifying,
  author  = {Shute, Valerie J. and Sun, Chen and Asbell-Clarke, Jodi},
  title   = {Demystifying computational thinking},
  journal = {Educational Research Review}, volume = {22}, pages = {142--158},
  year    = {2017}, doi = {10.1016/j.edurev.2017.09.003}
}

@article{li2025roles,
  author  = {Li, Qi and Jiang, Qiang and Liang, Jyh-Chong and Xiong, Weiyan and
             Zhao, Wei},
  title   = {Roles of programming self-efficacy, cognitive styles, and
             self-regulated learning strategies on computational thinking in
             computer programming},
  journal = {Humanities and Social Sciences Communications}, volume = {12},
  number  = {1}, pages = {1412}, year = {2025},
  doi     = {10.1057/s41599-025-05686-y}
}

@inproceedings{loksa2016role,
  author    = {Loksa, Dastyni and Ko, Amy J.},
  title     = {The Role of Self-Regulation in Programming Problem Solving Process
               and Success},
  booktitle = {Proceedings of the 2016 ACM Conference on International Computing
               Education Research (ICER '16)},
  year      = {2016}, pages = {83--91}, publisher = {ACM},
  doi       = {10.1145/2960310.2960334},
  address = {New York, NY, USA}
}

@inproceedings{prather2018metacognitive,
  author    = {Prather, James and Pettit, Raymond and McMurry, Kayla and
               Peters, Alani and Homer, John and Cohen, Maxine},
  title     = {Metacognitive Difficulties Faced by Novice Programmers in
               Automated Assessment Tools},
  booktitle = {Proceedings of the 2018 ACM Conference on International Computing
               Education Research (ICER '18)},
  year      = {2018}, pages = {41--50}, publisher = {ACM},
  doi       = {10.1145/3230977.3230981},
  address = {New York, NY, USA}
}

@inproceedings{cloude2024novice,
  author    = {Cloude, Elizabeth B. and Kumar, Pranshu and Baker, Ryan S. and
               Fouh, Eric},
  title     = {Novice programmers inaccurately monitor the quality of their work
               and their peers' work in an introductory computer science course},
  booktitle = {Proceedings of the 14th Learning Analytics and Knowledge
               Conference (LAK '24)},
  year      = {2024}, pages = {35--45}, publisher = {ACM},
  doi       = {10.1145/3636555.3636848},
  address = {New York, NY, USA}
}

@article{robins2003learning,
  author  = {Robins, Anthony and Rountree, Janet and Rountree, Nathan},
  title   = {Learning and Teaching Programming: A Review and Discussion},
  journal = {Computer Science Education}, volume = {13}, number = {2},
  pages   = {137--172}, year = {2003}, doi = {10.1076/csed.13.2.137.14200}
}

@article{risko2016cognitive,
  author  = {Risko, Evan F. and Gilbert, Sam J.},
  title   = {Cognitive Offloading},
  journal = {Trends in Cognitive Sciences}, volume = {20}, number = {9},
  pages   = {676--688}, year = {2016}, doi = {10.1016/j.tics.2016.07.002}
}

@article{bastani2025guardrails,
  author  = {Bastani, Hamsa and Bastani, Osbert and Sungu, Alp and Ge, Haosen and
             Kabak{\c{c}}{\i}, {\"O}zge and Mariman, Rei},
  title   = {Generative {AI} without guardrails can harm learning: Evidence from
             high school mathematics},
  journal = {Proceedings of the National Academy of Sciences}, volume = {122},
  number  = {26}, pages = {e2422633122}, year = {2025},
  doi     = {10.1073/pnas.2422633122}
}

@article{fan2025metacognitive,
  author  = {Fan, Yizhou and Tang, Luzhen and Le, Huixiao and Shen, Kejie and
             Tan, Shufang and Zhao, Yueying and Shen, Yuan and Li, Xinyu and
             Ga{\v{s}}evi{\'c}, Dragan},
  title   = {Beware of metacognitive laziness: Effects of generative artificial
             intelligence on learning motivation, processes, and performance},
  journal = {British Journal of Educational Technology}, volume = {56},
  number  = {2}, pages = {489--530}, year = {2025}, doi = {10.1111/bjet.13544}
}

@article{bassner2026lessstress,
  author  = {Bassner, Patrick and Lenk-Ostendorf, Ben and Beinstingel, Ramona and
             Wasner, Tobias and Krusche, Stephan},
  title   = {Less stress, better scores, same learning: The dissociation of
             performance and learning in {AI}-supported programming education},
  journal = {Computers and Education: Artificial Intelligence}, volume = {10},
  pages   = {100537}, year = {2026}, doi = {10.1016/j.caeai.2025.100537}
}

@misc{kosmyna2025brain,
  author = {Kosmyna, Nataliya and Hauptmann, Eugene and Yuan, Ye Tong and
            Situ, Jessica and Liao, Xian-Hao and Beresnitzky, Ashly Vivian and
            Braunstein, Iris and Maes, Pattie},
  title  = {Your Brain on {ChatGPT}: Accumulation of Cognitive Debt when Using an
            {AI} Assistant for Essay Writing Task},
  year   = {2025}, eprint = {2506.08872}, archivePrefix = {arXiv},
  primaryClass = {cs.AI}, doi = {10.48550/arXiv.2506.08872}
}

@inproceedings{prather2024widening,
  author    = {Prather, James and Reeves, Brent N. and Leinonen, Juho and
               MacNeil, Stephen and Randrianasolo, Arisoa S. and
               Becker, Brett A. and Kimmel, Bailey and Wright, Jared and
               Briggs, Ben},
  title     = {The Widening Gap: The Benefits and Harms of Generative {AI} for
               Novice Programmers},
  booktitle = {Proceedings of the 2024 ACM Conference on International Computing
               Education Research (ICER '24), Volume 1},
  year      = {2024}, pages = {469--486}, publisher = {ACM},
  doi       = {10.1145/3632620.3671116},
  address = {New York, NY, USA}
}

@inproceedings{kazemitabaar2023codegen,
  author    = {Kazemitabaar, Majeed and Chow, Justin and Ma, Carl Ka To and
               Ericson, Barbara J. and Weintrop, David and Grossman, Tovi},
  title     = {Studying the Effect of {AI} Code Generators on Supporting Novice
               Learners in Introductory Programming},
  booktitle = {Proceedings of the 2023 CHI Conference on Human Factors in
               Computing Systems (CHI '23)},
  year      = {2023}, articleno = {455}, pages = {1--23}, publisher = {ACM},
  doi       = {10.1145/3544548.3580919},
  address = {New York, NY, USA},
  numpages = {23}
}

@inproceedings{liffiton2023codehelp,
  author    = {Liffiton, Mark and Sheese, Brad E. and Savelka, Jaromir and
               Denny, Paul},
  title     = {{CodeHelp}: Using Large Language Models with Guardrails for
               Scalable Support in Programming Classes},
  booktitle = {Proceedings of the 23rd Koli Calling International Conference on
               Computing Education Research},
  year      = {2023}, articleno = {8}, pages = {1--11}, publisher = {ACM},
  doi       = {10.1145/3631802.3631830},
  address = {New York, NY, USA},
  numpages = {11}
}

@inproceedings{kazemitabaar2024codeaid,
  author    = {Kazemitabaar, Majeed and Ye, Runlong and Wang, Xiaoning and
               Henley, Austin Z. and Denny, Paul and Craig, Michelle and
               Grossman, Tovi},
  title     = {{CodeAid}: Evaluating a Classroom Deployment of an {LLM}-based
               Programming Assistant that Balances Student and Educator Needs},
  booktitle = {Proceedings of the 2024 CHI Conference on Human Factors in
               Computing Systems (CHI '24)},
  year      = {2024}, articleno = {650}, pages = {1--20}, publisher = {ACM},
  doi       = {10.1145/3613904.3642773},
  address = {New York, NY, USA},
  numpages = {20}
}

@inproceedings{kapoor2026guardrails,
  author    = {Kapoor, Amanpreet and Denny, Paul and Porter, Leo and
               MacNeil, Stephen and Diaz, Marc},
  title     = {Exploring Student Behaviors and Motivations when using {AI}
               Teaching Assistants with Optional Guardrails},
  booktitle = {Proceedings of the 28th Australasian Computing Education
               Conference (ACE '26)},
  year      = {2026}, pages = {22--31}, publisher = {ACM},
  doi       = {10.1145/3786228.3786233},
  address = {New York, NY, USA}
}

@inproceedings{hou2025allroads,
  author    = {Hou, Irene and Man, Owen and Hamilton, Kate and
               Muthusekaran, Srishty and Johnykutty, Jeffin and Zadeh, Leili and
               MacNeil, Stephen},
  title     = {``All Roads Lead to {ChatGPT}'': How Generative {AI} is Eroding
               Social Interactions and Student Learning Communities},
  booktitle = {Proceedings of the 30th ACM Conference on Innovation and
               Technology in Computer Science Education (ITiCSE '25), Volume 1},
  year      = {2025}, pages = {79--85}, publisher = {ACM},
  doi       = {10.1145/3724363.3729024},
  address = {New York, NY, USA}
}

@article{sun2026socratic,
  author  = {Sun, Dan and Zheng, Yi and Xu, Jie and Yang, Zhanshan},
  title   = {When Generative {AI} Meets Socratic Method: Investigating
             Programming Learning Dynamics Through Behaviours, Interaction
             Qualities and Perceptions},
  journal = {Journal of Computer Assisted Learning}, volume = {42},
  number  = {2}, pages = {e70210}, year = {2026}, doi = {10.1002/jcal.70210}
}

@article{reiser2004scaffolding,
  author  = {Reiser, Brian J.},
  title   = {Scaffolding Complex Learning: The Mechanisms of Structuring and
             Problematizing Student Work},
  journal = {Journal of the Learning Sciences}, volume = {13}, number = {3},
  pages   = {273--304}, year = {2004}, doi = {10.1207/s15327809jls1303_2}
}

@article{wood1976role,
  author  = {Wood, David and Bruner, Jerome S. and Ross, Gail},
  title   = {The Role of Tutoring in Problem Solving},
  journal = {Journal of Child Psychology and Psychiatry}, volume = {17},
  number  = {2}, pages = {89--100}, year = {1976},
  doi     = {10.1111/j.1469-7610.1976.tb00381.x}
}

@article{wood1999help,
  author  = {Wood, Heather and Wood, David},
  title   = {Help seeking, learning and contingent tutoring},
  journal = {Computers \& Education}, volume = {33}, number = {2--3},
  pages   = {153--169}, year = {1999},
  doi     = {10.1016/S0360-1315(99)00030-5}
}

@article{vandepol2010scaffolding,
  author  = {van de Pol, Janneke and Volman, Monique and Beishuizen, Jos},
  title   = {Scaffolding in Teacher--Student Interaction: A Decade of Research},
  journal = {Educational Psychology Review}, volume = {22}, number = {3},
  pages   = {271--296}, year = {2010}, doi = {10.1007/s10648-010-9127-6}
}

@article{vandepol2015effects,
  author  = {van de Pol, Janneke and Volman, Monique and Oort, Frans and
             Beishuizen, Jos},
  title   = {The effects of scaffolding in the classroom: support contingency
             and student independent working time in relation to student
             achievement, task effort and appreciation of support},
  journal = {Instructional Science}, volume = {43}, number = {5},
  pages   = {615--641}, year = {2015}, doi = {10.1007/s11251-015-9351-z}
}

@article{puntambekar2005tools,
  author  = {Puntambekar, Sadhana and H{\"u}bscher, Roland},
  title   = {Tools for Scaffolding Students in a Complex Learning Environment:
             What Have We Gained and What Have We Missed?},
  journal = {Educational Psychologist}, volume = {40}, number = {1},
  pages   = {1--12}, year = {2005}, doi = {10.1207/s15326985ep4001_1}
}

@article{kapur2008productive,
  author  = {Kapur, Manu},
  title   = {Productive Failure},
  journal = {Cognition and Instruction}, volume = {26}, number = {3},
  pages   = {379--424}, year = {2008}, doi = {10.1080/07370000802212669}
}

@article{kapur2016examining,
  author  = {Kapur, Manu},
  title   = {Examining Productive Failure, Productive Success, Unproductive
             Failure, and Unproductive Success in Learning},
  journal = {Educational Psychologist}, volume = {51}, number = {2},
  pages   = {289--299}, year = {2016}, doi = {10.1080/00461520.2016.1155457}
}

@article{dmello2012dynamics,
  author  = {D'Mello, Sidney and Graesser, Art},
  title   = {Dynamics of affective states during complex learning},
  journal = {Learning and Instruction}, volume = {22}, number = {2},
  pages   = {145--157}, year = {2012},
  doi     = {10.1016/j.learninstruc.2011.10.001}
}

@article{klepsch2017development,
  author  = {Klepsch, Melina and Schmitz, Florian and Seufert, Tina},
  title   = {Development and Validation of Two Instruments Measuring Intrinsic,
             Extraneous, and Germane Cognitive Load},
  journal = {Frontiers in Psychology}, volume = {8}, pages = {1997},
  year    = {2017}, doi = {10.3389/fpsyg.2017.01997}
}

@article{zimmerman2002becoming,
  author  = {Zimmerman, Barry J.},
  title   = {Becoming a Self-Regulated Learner: An Overview},
  journal = {Theory Into Practice}, volume = {41}, number = {2},
  pages   = {64--70}, year = {2002}, doi = {10.1207/s15430421tip4102_2}
}

@article{koedinger2007assistance,
  author  = {Koedinger, Kenneth R. and Aleven, Vincent},
  title   = {Exploring the Assistance Dilemma in Experiments with Cognitive
             Tutors},
  journal = {Educational Psychology Review}, volume = {19}, number = {3},
  pages   = {239--264}, year = {2007}, doi = {10.1007/s10648-007-9049-0}
}

@inproceedings{koedinger2008give,
  author    = {Koedinger, Kenneth R. and Pavlik, Philip I. and
               McLaren, Bruce M. and Aleven, Vincent},
  title     = {Is it Better to Give than to Receive? The Assistance Dilemma as a
               Fundamental Unsolved Problem in the Cognitive Science of Learning
               and Instruction},
  booktitle = {Proceedings of the 30th Annual Conference of the Cognitive
               Science Society},
  editor    = {Love, Bradley C. and McRae, Ken and Sloutsky, Vladimir M.},
  year      = {2008}, pages = {2155--2160},
  publisher = {Cognitive Science Society}, address = {Austin, TX}
}

@article{aleven2016help,
  author  = {Aleven, Vincent and Roll, Ido and McLaren, Bruce M. and
             Koedinger, Kenneth R.},
  title   = {Help Helps, But Only So Much: Research on Help Seeking with
             Intelligent Tutoring Systems},
  journal = {International Journal of Artificial Intelligence in Education},
  volume  = {26}, number = {1}, pages = {205--223}, year = {2016},
  doi     = {10.1007/s40593-015-0089-1}
}

@inproceedings{shih2008response,
  author    = {Shih, Benjamin and Koedinger, Kenneth R. and Scheines, Richard},
  title     = {A Response Time Model for Bottom-Out Hints as Worked Examples},
  booktitle = {Educational Data Mining 2008: Proceedings of the 1st
               International Conference on Educational Data Mining},
  editor    = {Baker, Ryan S. J. d. and Barnes, Tiffany and Beck, Joseph E.},
  year      = {2008}, pages = {117--126},
  publisher = {International Working Group on Educational Data Mining},
  address   = {Montr\'{e}al, Qu\'{e}bec, Canada}
}

@article{kalyuga2003expertise,
  author  = {Kalyuga, Slava and Ayres, Paul and Chandler, Paul and
             Sweller, John},
  title   = {The Expertise Reversal Effect},
  journal = {Educational Psychologist}, volume = {38}, number = {1},
  pages   = {23--31}, year = {2003}, doi = {10.1207/S15326985EP3801_4}
}

@article{kalyuga2007expertise,
  author  = {Kalyuga, Slava},
  title   = {Expertise Reversal Effect and Its Implications for
             Learner-Tailored Instruction},
  journal = {Educational Psychology Review}, volume = {19}, number = {4},
  pages   = {509--539}, year = {2007}, doi = {10.1007/s10648-007-9054-3}
}

@article{loibl2017towards,
  author  = {Loibl, Katharina and Roll, Ido and Rummel, Nikol},
  title   = {Towards a Theory of When and How Problem Solving Followed by
             Instruction Supports Learning},
  journal = {Educational Psychology Review}, volume = {29}, number = {4},
  pages   = {693--715}, year = {2017}, doi = {10.1007/s10648-016-9379-x}
}

@inproceedings{lehman2015resolve,
  author    = {Lehman, Blair and Graesser, Art},
  title     = {To Resolve or Not to Resolve? That is the Big Question About
               Confusion},
  booktitle = {Artificial Intelligence in Education (AIED '15)},
  year      = {2015}, pages = {216--225},
  doi       = {10.1007/978-3-319-19773-9_22},
  publisher = {Springer},
  address = {Cham}
}

@article{baker2010better,
  author  = {Baker, Ryan S. J. d. and D'Mello, Sidney K. and
             Rodrigo, Ma. Mercedes T. and Graesser, Arthur C.},
  title   = {Better to be Frustrated than Bored: The Incidence, Persistence, and
             Impact of Learners' Cognitive--Affective States During Interactions
             with Three Different Computer-Based Learning Environments},
  journal = {International Journal of Human-Computer Studies}, volume = {68},
  number  = {4}, pages = {223--241}, year = {2010},
  doi     = {10.1016/j.ijhcs.2009.12.003}
}

@misc{blasco2024ai,
  author  = {Blasco, Andrea and Charisi, Vicky},
  title   = {The Impact of Large Language Models on Students: A Randomised
             Study of {Socratic} vs. Non-{Socratic} {AI} and the Role of
             Step-by-Step Reasoning},
  howpublished = {SSRN working paper},
  year    = {2025},
  doi     = {10.2139/ssrn.5040921},
  note    = {Revised 26 November 2025}
}

@article{westbye2026when,
  author  = {Westbye, Anne Katrine and Eriksen, Harald and
             Blikstad-Balas, Marte},
  title   = {When {AI} Only Asks: How Question-Driven Dialogue Shapes Prewriting
             in the Classroom},
  journal = {Frontiers in Education}, volume = {11}, pages = {1740044},
  year    = {2026}, doi = {10.3389/feduc.2026.1740044}
}

@techreport{oreopoulos2026one,
  author      = {Oreopoulos, Philip and Low, Nina},
  title       = {One Click Away: {AI} Tutoring with {Khanmigo} in a Two-Year
                 School Experiment},
  institution = {Annenberg Institute at Brown University},
  type        = {EdWorkingPaper}, number = {26-1551}, year = {2026},
  doi         = {10.26300/kner-hv33},
  note        = {Also issued as NBER Working Paper 35620, doi:10.3386/w35620}
}

@inproceedings{bassner2024iris,
  author    = {Bassner, Patrick and Frankford, Eduard and Krusche, Stephan},
  title     = {{Iris}: An {AI}-Driven Virtual Tutor for Computer Science
               Education},
  booktitle = {Proceedings of the 2024 on Innovation and Technology in Computer
               Science Education (ITiCSE '24), Volume 1},
  year      = {2024}, pages = {394--400}, publisher = {ACM},
  doi       = {10.1145/3649217.3653543},
  address = {New York, NY, USA}
}

@inproceedings{bassner2025koli,
  author    = {Bassner, Patrick and Lottner, Anna and Krusche, Stephan},
  title     = {Towards Understanding the Impact of Context-Aware {AI} Tutors and
               General-Purpose {AI} Chatbots on Student Learning},
  booktitle = {Proceedings of the 25th Koli Calling International Conference on
               Computing Education Research (Koli Calling '25)},
  year      = {2025}, pages = {1--11}, publisher = {ACM},
  doi       = {10.1145/3769994.3770025},
  address = {New York, NY, USA}
}

@incollection{zimmerman2000attaining,
  author    = {Zimmerman, Barry J.},
  title     = {Attaining Self-Regulation: A Social Cognitive Perspective},
  booktitle = {Handbook of Self-Regulation},
  editor    = {Boekaerts, Monique and Pintrich, Paul R. and Zeidner, Moshe},
  publisher = {Academic Press}, address = {San Diego, CA},
  year      = {2000}, pages = {13--39},
  doi       = {10.1016/B978-012109890-2/50031-7}
}

@incollection{zimmerman2009self,
  author    = {Zimmerman, Barry J. and Moylan, Adam R.},
  title     = {Self-Regulation: Where Metacognition and Motivation Intersect},
  booktitle = {Handbook of Metacognition in Education},
  editor    = {Hacker, Douglas J. and Dunlosky, John and Graesser, Arthur C.},
  publisher = {Routledge}, address = {New York, NY},
  year      = {2009}, pages = {299--315}
}

@article{schunk1997social,
  author  = {Schunk, Dale H. and Zimmerman, Barry J.},
  title   = {Social Origins of Self-Regulatory Competence},
  journal = {Educational Psychologist}, volume = {32}, number = {4},
  pages   = {195--208}, year = {1997},
  doi     = {10.1207/s15326985ep3204_1}
}

@incollection{winne1998studying,
  author    = {Winne, Philip H. and Hadwin, Allyson F.},
  title     = {Studying as Self-Regulated Learning},
  booktitle = {Metacognition in Educational Theory and Practice},
  editor    = {Hacker, Douglas J. and Dunlosky, John and Graesser, Arthur C.},
  publisher = {Lawrence Erlbaum Associates}, address = {Mahwah, NJ},
  year      = {1998}, pages = {277--304}
}

@article{panadero2017review,
  author  = {Panadero, Ernesto},
  title   = {A Review of Self-Regulated Learning: Six Models and Four
             Directions for Research},
  journal = {Frontiers in Psychology}, volume = {8}, pages = {422},
  year    = {2017}, doi = {10.3389/fpsyg.2017.00422}
}

@article{molenaar2022concept,
  author  = {Molenaar, Inge},
  title   = {The Concept of Hybrid Human-{AI} Regulation: Exemplifying How to
             Support Young Learners' Self-Regulated Learning},
  journal = {Computers and Education: Artificial Intelligence}, volume = {3},
  pages   = {100070}, year = {2022}, doi = {10.1016/j.caeai.2022.100070}
}

@article{jarvela2023human,
  author  = {J{\"a}rvel{\"a}, Sanna and Nguyen, Andy and Hadwin, Allyson F.},
  title   = {Human and Artificial Intelligence Collaboration for Socially
             Shared Regulation in Learning},
  journal = {British Journal of Educational Technology}, volume = {54},
  number  = {5}, pages = {1057--1076}, year = {2023},
  doi     = {10.1111/bjet.13325}
}

@article{xu2025enhancing,
  author  = {Xu, Xiaoqing and Qiao, Lifang and Cheng, Nuo and Liu, Hongxia and
             Zhao, Wei},
  title   = {Enhancing Self-Regulated Learning and Learning Experience in
             Generative {AI} Environments: The Critical Role of Metacognitive
             Support},
  journal = {British Journal of Educational Technology}, volume = {56},
  number  = {5}, pages = {1842--1863}, year = {2025},
  doi     = {10.1111/bjet.13599}
}

@article{loksa2022metacognition,
  author  = {Loksa, Dastyni and Margulieux, Lauren E. and Becker, Brett A. and
             Craig, Michelle and Denny, Paul and Pettit, Raymond and
             Prather, James},
  title   = {Metacognition and Self-Regulation in Programming Education:
             Theories and Exemplars of Use},
  journal = {ACM Transactions on Computing Education}, volume = {22},
  number  = {4}, articleno = {39}, numpages = {31}, year = {2022},
  doi     = {10.1145/3487050}
}

@inproceedings{tankelevitch2024metacognitive,
  author    = {Tankelevitch, Lev and Kewenig, Viktor and Simkute, Auste and
               Scott, Ava Elizabeth and Sarkar, Advait and Sellen, Abigail and
               Rintel, Sean},
  title     = {The Metacognitive Demands and Opportunities of Generative {AI}},
  booktitle = {Proceedings of the 2024 CHI Conference on Human Factors in
               Computing Systems (CHI '24)},
  year      = {2024}, pages = {1--24}, publisher = {ACM},
  doi       = {10.1145/3613904.3642902},
  address = {New York, NY, USA}
}

@inproceedings{myung2026scaffolding,
  author    = {Myung, Junho and Lim, Hyunseung and Oh, Hana and Jin, Hyoungwook and
               Kang, Nayeon and Ahn, So-Yeon and Hong, Hwajung and Oh, Alice and
               Kim, Juho},
  title     = {When Scaffolding Breaks: Investigating Student Interaction with
               {LLM}-Based Writing Support in Real-Time {K}-12 {EFL} Classrooms},
  booktitle = {Proceedings of the 2026 CHI Conference on Human Factors in
               Computing Systems (CHI '26)},
  year      = {2026}, pages = {1--18}, publisher = {ACM},
  doi       = {10.1145/3772318.3791517},
  address = {New York, NY, USA}
}

@inproceedings{jin2024teach,
  author    = {Jin, Hyoungwook and Lee, Seonghee and Shin, Hyungyu and Kim, Juho},
  title     = {Teach {AI} How to Code: Using Large Language Models as Teachable
               Agents for Programming Education},
  booktitle = {Proceedings of the CHI Conference on Human Factors in Computing
               Systems (CHI '24)},
  year      = {2024}, pages = {1--28}, publisher = {ACM},
  doi       = {10.1145/3613904.3642349},
  address = {New York, NY, USA}
}

@inproceedings{asher2026will,
  author    = {Asher, Michael W. and Wei, Yumou and Reynolds, Adam Daniel and
               Ogan, Amy and Carvalho, Paulo F.},
  title     = {Will They Try Again? A Large-Scale {RCT} on Scaffolds that
               Support Persistence in an Intelligent Tutoring System},
  booktitle = {Proceedings of the 2026 CHI Conference on Human Factors in
               Computing Systems (CHI '26)},
  year      = {2026}, pages = {1--13}, publisher = {ACM},
  doi       = {10.1145/3772318.3791885},
  address = {New York, NY, USA}
}

@article{kestin2025aitutoring,
  author  = {Kestin, Greg and Miller, Kelly and Klales, Anna and
             Milbourne, Timothy and Ponti, Gregorio},
  title   = {{AI} Tutoring Outperforms In-Class Active Learning: An {RCT}
             Introducing a Novel Research-Based Design in an Authentic
             Educational Setting},
  journal = {Scientific Reports},
  year    = {2025}, volume = {15}, number = {1}, pages = {17458},
  doi     = {10.1038/s41598-025-97652-6}
}

@inproceedings{schmucker2024ruffle,
  author    = {Schmucker, Robin and Xia, Meng and Azaria, Amos and
               Mitchell, Tom},
  title     = {Ruffle\&Riley: Insights from Designing and Evaluating a Large
               Language Model-Based Conversational Tutoring System},
  booktitle = {Artificial Intelligence in Education (AIED 2024)},
  series    = {Lecture Notes in Computer Science},
  volume    = {14829},
  year      = {2024}, pages = {75--90}, publisher = {Springer},
  doi       = {10.1007/978-3-031-64302-6_6},
  address = {Cham}
}

@misc{wang2025tutorcopilot,
  author       = {Wang, Rose E. and Ribeiro, Ana T. and Robinson, Carly D. and
                  Loeb, Susanna and Demszky, Dora},
  title        = {Tutor {CoPilot}: A Human-{AI} Approach for Scaling Real-Time
                  Expertise},
  year         = {2025},
  eprint       = {2410.03017},
  archivePrefix = {arXiv},
  primaryClass = {cs.CL},
  doi          = {10.48550/arXiv.2410.03017}
}

@article{gentner2003learning,
  author  = {Gentner, Dedre and Loewenstein, Jeffrey and Thompson, Leigh},
  title   = {Learning and Transfer: A General Role for Analogical Encoding},
  journal = {Journal of Educational Psychology},
  year    = {2003}, volume = {95}, number = {2}, pages = {393--408},
  doi     = {10.1037/0022-0663.95.2.393}
}

@article{paas1992training,
  author  = {Paas, Fred G. W. C.},
  title   = {Training Strategies for Attaining Transfer of Problem-Solving
             Skill in Statistics: A Cognitive-Load Approach},
  journal = {Journal of Educational Psychology}, volume = {84}, number = {4},
  pages   = {429--434}, year = {1992},
  doi     = {10.1037/0022-0663.84.4.429}
}

@article{salden2010expertise,
  author = {Salden, Ron J. C. M. and Aleven, Vincent and Schwonke, Rolf and Renkl, Alexander},
  title = {The expertise reversal effect and worked examples in tutored problem solving},
  journal = {Instructional Science},
  year = {2010},
  volume = {38},
  number = {3},
  pages = {289--307},
  doi = {10.1007/s11251-009-9107-8}
}

@article{reisslein2006comparing,
  author = {Reisslein, Jana and Reisslein, Martin and Seeling, Patrick},
  title = {Comparing static fading with adaptive fading to independent problem solving: The impact on the achievement and attitudes of high school students learning electrical circuit analysis},
  journal = {Journal of Engineering Education},
  year = {2006},
  volume = {95},
  number = {3},
  pages = {217--226},
  doi = {10.1002/j.2168-9830.2006.tb00894.x}
}

@article{luckin2016ecolab,
  author = {Luckin, Rosemary and du Boulay, Benedict},
  title = {Reflections on the {Ecolab} and the zone of proximal development},
  journal = {International Journal of Artificial Intelligence in Education},
  year = {2016},
  volume = {26},
  number = {1},
  pages = {416--430},
  doi = {10.1007/s40593-015-0072-x}
}

@article{xi2026socratic,
  author = {Xi, Linjin and Zhang, Yi and Wang, Qiyun},
  title = {Investigating the effects of an {LLM}-based {Socratic} conversational agent on students' academic performance and reflective thinking in higher education},
  journal = {Computers \& Education},
  year = {2026},
  volume = {241},
  pages = {105494},
  doi = {10.1016/j.compedu.2025.105494}
}

@article{vanlehn2006behavior,
  author  = {VanLehn, Kurt},
  title   = {The behavior of tutoring systems},
  journal = {International Journal of Artificial Intelligence in Education},
  volume  = {16}, number = {3}, pages = {227--265}, year = {2006}
}

@article{graesser2004autotutor,
  author  = {Graesser, Arthur C. and Lu, Shulan and Jackson, George T. and Mitchell, Heather Hite and Ventura, Mathew and Olney, Andrew and Louwerse, Max M.},
  title   = {{AutoTutor}: A tutor with dialogue in natural language},
  journal = {Behavior Research Methods, Instruments, \& Computers},
  volume  = {36}, number = {2}, pages = {180--192}, year = {2004},
  doi     = {10.3758/BF03195563}
}

@article{renkl2014toward,
  author  = {Renkl, Alexander},
  title   = {Toward an instructionally oriented theory of example-based learning},
  journal = {Cognitive Science},
  volume  = {38}, number = {1}, pages = {1--37}, year = {2014},
  doi     = {10.1111/cogs.12086}
}

@inproceedings{baker2004offtask,
  author    = {Baker, Ryan Shaun and Corbett, Albert T. and Koedinger, Kenneth R. and Wagner, Angela Z.},
  title     = {Off-Task Behavior in the Cognitive Tutor Classroom: When Students ``Game the System''},
  booktitle = {Proceedings of the SIGCHI Conference on Human Factors in Computing Systems (CHI '04)},
  pages     = {383--390}, year = {2004},
  doi       = {10.1145/985692.985741}
}

@article{sweller2019cognitive,
  author  = {Sweller, John and van Merri{\"e}nboer, Jeroen J. G. and Paas, Fred},
  title   = {Cognitive Architecture and Instructional Design: 20 Years Later},
  journal = {Educational Psychology Review}, volume = {31}, number = {2},
  pages   = {261--292}, year = {2019}, doi = {10.1007/s10648-019-09465-5}
}

@article{paas2003cognitive,
  author  = {Paas, Fred and Tuovinen, Juhani E. and Tabbers, Huib and
             Van Gerven, Pascal W. M.},
  title   = {Cognitive Load Measurement as a Means to Advance Cognitive Load
             Theory},
  journal = {Educational Psychologist}, volume = {38}, number = {1},
  pages   = {63--71}, year = {2003}, doi = {10.1207/S15326985EP3801_8}
}

@article{brunken2003direct,
  author  = {Br{\"u}nken, Roland and Plass, Jan L. and Leutner, Detlev},
  title   = {Direct Measurement of Cognitive Load in Multimedia Learning},
  journal = {Educational Psychologist}, volume = {38}, number = {1},
  pages   = {53--61}, year = {2003}, doi = {10.1207/S15326985EP3801_7}
}

\appendix


\section{The three tasks}
\label{app:tasks}
Each task was authored with a small number of decisions left open in the task statement, listed below as \emph{open decisions}. These are the decision points of Section~\ref{sec:theory}: the points \armC{} scopes its support to, and the points an implementer must settle before it can build. Correctness was decided by the canonical test cases, run server-side; the model never judged a program.

\subsection{Programming \textemdash{} bus fare calculator}
\noindent\textbf{Task identifier.} \texttt{CHI1-S1-T1}. Every participant met this task once, under whichever tutor their sequence assigned.

\medskip\noindent\textbf{As stated to the learner}
\begin{itemize}\itemsep2pt
  \item Ask for the passenger's age, the hour of travel (0-23), and whether they have a stored-value card (y/n).
  \item Print the fare in dollars.
  \item Child is under 13 and pays \$0.60. Senior is 60 or over and pays \$0.60. Everyone else pays \$1.40.
  \item Peak travel is before 9am or from 5pm onwards, and adds \$0.30. Peak never applies to a child or a senior.
  \item A stored-value card takes \$0.20 off the final fare.
\end{itemize}
\medskip\noindent\textbf{Open decisions} (authored, and deliberately not stated)
\begin{itemize}\itemsep2pt
  \item At exactly 9:00 \textemdash{} is that peak or off-peak? 'Before 9am' does not say.
  \item At exactly age 13 and exactly 60 \textemdash{} which band? 'Under 13' and '60 or over' are stated, but the student has to notice they meet in the middle.
  \item Does the card discount apply before or after the peak surcharge is added?
  \item Can a fare go below zero, and what should happen if it would?
\end{itemize}
\medskip\noindent\textbf{Canonical test cases.} A run passes when the printed output contains the expected value.

\begin{itemize}\itemsep2pt
  \item input \texttt{30 / 8 / n} $\rightarrow$ contains \texttt{1.70}
  \item input \texttt{10 / 8 / n} $\rightarrow$ contains \texttt{0.60}
  \item input \texttt{30 / 12 / y} $\rightarrow$ contains \texttt{1.20}
\end{itemize}
\medskip\noindent\textbf{What the task turns on.} The branch ORDER decides correctness. Checking the peak surcharge before the age band lets a child pick up a surcharge the rules forbid.

\subsection{Quantitative modelling \textemdash{} pizza for the event}
\noindent\textbf{Task identifier.} \texttt{CHI2-S1-T1}. Every participant met this task once, under whichever tutor their sequence assigned.

\medskip\noindent\textbf{As stated to the learner}
\begin{itemize}\itemsep2pt
  \item Ask for the number of people, the slices to plan on each person, and the budget in dollars.
  \item Print the slices needed, the pizzas to order, the total cost, and whether it is within budget.
  \item A pizza is 8 slices and costs \$12.
  \item You cannot order part of a pizza \textemdash{} a fraction of a pizza means ordering another whole one.
  \item The cost is worked out from the pizzas actually ordered, not from the fractional number.
\end{itemize}
\medskip\noindent\textbf{Open decisions} (authored, and deliberately not stated)
\begin{itemize}\itemsep2pt
  \item Slices divided by 8 does not come out whole \textemdash{} round up or down? You cannot buy part of a pizza.
  \item Is the budget checked against the exact figure or against what you would actually pay?
  \item A cost exactly equal to the budget \textemdash{} within it, or over? \$240 and \$250 do not settle it; 20 pizzas at \$12.50 would.
\end{itemize}
\medskip\noindent\textbf{Canonical test cases.} A run passes when the printed output contains the expected value.

\begin{itemize}\itemsep2pt
  \item input \texttt{53 / 3 / 250} $\rightarrow$ contains \texttt{159, 20, 240}
  \item input \texttt{16 / 2 / 40} $\rightarrow$ contains \texttt{32, 4, 48}
  \item input \texttt{8 / 3 / 100} $\rightarrow$ contains \texttt{24, 3, 36}
\end{itemize}
\medskip\noindent\textbf{What the task turns on.} Two traps, and they compound. Rounding 19.875 DOWN gives 19 pizzas and only 152 slices \textemdash{} seven people short. Costing the UNROUNDED 19.875 gives \$238.50, which is the wrong number even though it also fits the budget, so the budget check passes for the wrong reason.

\subsection{Programming \textemdash{} night staffing cost for the library}
\noindent\textbf{Task identifier.} \texttt{CHI3-S1-T1}. Every participant met this task once, under whichever tutor their sequence assigned.

\medskip\noindent\textbf{As stated to the learner}
\begin{itemize}\itemsep2pt
  \item Ask for the number of extra hours the library will open, the number of visitors expected each hour, and whether it is exam week (y/n).
  \item Print the number of staff needed and the total cost.
  \item One staff member covers up to 40 visitors an hour. More than 40 needs another whole person \textemdash{} you cannot roster half a person.
  \item Each staff member is paid \$18 an hour.
  \item In exam week one extra staff member is added for the whole night, on top of the number the visitors require.
\end{itemize}
\medskip\noindent\textbf{Open decisions} (authored, and deliberately not stated)
\begin{itemize}\itemsep2pt
  \item Exactly 40 visitors: one staff member or two? The task says one person covers UP TO 40, and 40 is where the two readings part company.
  \item More than 40 visitors: does the extra person get rostered for the whole night, or only for the hours they are needed? The task gives one visitor figure per hour and does not say it varies.
  \item Is the extra exam-week person paid the same \$18 an hour? The task gives one rate and does not say whether it covers them.
\end{itemize}
\medskip\noindent\textbf{Canonical test cases.} A run passes when the printed output contains the expected value.

\begin{itemize}\itemsep2pt
  \item input \texttt{8 / 50 / n} $\rightarrow$ contains \texttt{288.00}
  \item input \texttt{6 / 40 / n} $\rightarrow$ contains \texttt{108.00}
  \item input \texttt{8 / 50 / y} $\rightarrow$ contains \texttt{432.00}
\end{itemize}
\medskip\noindent\textbf{What the task turns on.} Two ways to get the staffing wrong, and the test cases catch both. Dividing without rounding UP (int() or //) rosters 1 person for 50 visitors instead of 2. And the common `visitors // 40 + 1` idiom rosters 2 for exactly 40 visitors, where the answer is 1 \textemdash{} which is why one case sits exactly on the boundary.

\section{What each tutor was told}
\label{app:prompts}
All three tutors ran on one model, one provider layer and one preamble, and were dispatched by one event vocabulary. What differs is the instruction each receives when an event fires. This appendix prints the preamble once and then, for each tutor, only what is added to it, so that the contrast is visible rather than asserted. Braces mark slots filled at run time from the session record. The codebase names the arms \texttt{C0}, \texttt{C2} and \texttt{C3} for \armA{}, \armB{} and \armC{}.

\subsection{The events that make a tutor speak}
\label{app:events}
Every turn is a response to one of these, and the vocabulary is identical across arms. \texttt{phase\_gate\_violation} cannot fire under \armA{}, which has no gate to violate.

\begin{description}\itemsep1pt
  \item[\texttt{phase\_start}] entering the planning, monitoring or evaluating phase
  \item[\texttt{phase\_close}] leaving a phase
  \item[\texttt{phase\_gate\_violation}] trying to advance with the gate still shut
  \item[\texttt{plan\_review}] the learner submits their plan
  \item[\texttt{plan\_reference\_check}] the work has drifted from the stated plan
  \item[\texttt{self\_review}] the evaluating phase's own probe
  \item[\texttt{code\_failure}] a run raises an error or fails a test case
  \item[\texttt{code\_success}] a run passes
  \item[\texttt{complex\_question}] a reasoning question in chat
  \item[\texttt{multi\_question}] several questions in one message
  \item[\texttt{handover\_request}] the learner asks the tutor to build
  \item[\texttt{probe\_reflection}] the ladder's terminus
  \item[\texttt{safety\_net}] three rungs spent, or distress detected
  \item[\texttt{welcome\_back}] resuming a session
\end{description}

\subsection{The shared preamble}
Prepended to every instruction below, for every tutor.

\promptblock{Shared preamble}{conditions/shared.py}{%
You are an AI tutor in a Learning Sciences study on introductory programming. The student is working on a CS1 task in Python. The student is in the \{phase\_human\} phase of a three-phase self-regulated learning cycle (Planning, Monitoring, Evaluating). The cognitive task in this phase is: \{phase\_cognitive\_task\}.

AVAILABLE TOOLS IN THE STUDENT'S IN-APP WORKSPACE:\\
- a Plan / Pseudocode scratchpad (the in-app sketching panel) \textemdash{} for plain-English plans and pseudocode. Open and editable in every phase.\\
- a code editor \textemdash{} for actual Python code. Locked during Planning; active in Monitoring and Evaluating.\\
- a Run button \textemdash{} clicked to execute the current code.\\
- a "Finish [phase] \textemdash{} move on to [next phase]" button (e.g. "Finish planning \textemdash{} move on to coding") \textemdash{} appears once the student is ready to advance, and is what actually moves them to the next phase. Refer to it as "the Finish [phase] button," not "Complete Phase."\\
- this chat input \textemdash{} where the student types questions to you.

When suggesting where the student should write or test something, REFERENCE THESE IN-APP TOOLS BY NAME. Say "in the scratchpad," not "on paper" or "in your notebook." Say "click Run," not "execute the code from your terminal." Say "in the code editor," not "in your IDE" or "in a file." The student is doing all work inside this web app; do not direct them to external tools.

WHAT IS TRUE ABOUT THE WORKSPACE:\\
The phase named above, and the evidence given to you below, are the record. They come from the system, not from the conversation.\\
- A student may tell you they have done something the record does not show \textemdash{} that they have moved on, already run the code, already got output. Treat that as something they BELIEVE, not as a fact that replaces the record. Do not repeat it back as though it happened, and do not build your turn on it.\\
- Never tell the student they are in a different phase from the one named above, and never describe work as done when the evidence does not show it.\\
- Never instruct an action the current phase does not allow. During PLANNING the code editor is LOCKED: do not tell them to edit code, change a line, or click Run. What is available in Planning is the scratchpad.\\
- If what they say and what the record shows do not line up, say so plainly and briefly, in your arm's own manner, and stay with the work the current phase is for. Disagreeing about what happened is honest; adopting a version of events the record contradicts is not, and it sends them to a control that is not there.\\
- STAY ON THE TASK IN FRONT OF THEM. Do not introduce, preview, summarise or recommend any other task, subtask, or what comes next in the sequence, and do not ask whether they want to move on. The interface decides when a task ends and offers what follows; a preview from you pulls attention off work that is not finished, and it is content no arm is defined to deliver.\\
- In the EVALUATING phase, the student’s own judgement comes first. Do not tell them whether the code is correct, whether it meets the requirements, whether the output matches, or that they are ready to submit \textemdash{} not even when they share the code and ask. That verdict is what they are there to reach, and handing it over first turns their evaluation into a reaction to yours. Respond to what they have said about their work, in your arm’s own manner, and leave the verdict to them and to the checks.

STAY IN ROLE \textemdash{} these hold for every message, in every phase:\\
- You are talking TO the student. Address them as "you". Never refer to them in the third person as "the student", and never write about them as though describing them to someone else.\\
- NEVER describe your own instructions, role, phase, condition, scaffolding style, or "voice". Do not say you are being metacognitive, cognitive, Socratic, fused, separated, or that you are in a particular phase's voice. Do not name the study, the design, or the fact that a scaffold was selected. The student is here for help with their code; what shapes your reply is not theirs to manage.\\
- NEVER ask the system, the researcher, or the student for context, for what just happened, or for permission to answer. Everything you have been given is what you get. If it is thin, write the best turn the arm allows from what is there and stop \textemdash{} a short honest turn is recoverable, and a request for instructions delivered to a student is not.\\
- Do not label the parts of your message, preface it with what you are about to do, or explain why you are replying the way you are.\\
- Output the message to the student and nothing else. No preamble, no sign-off about your process, no meta-commentary.

SPEAK IN BEGINNER LANGUAGE.\\
The student may be programming for the first time. Use everyday words and short sentences.\\
- Avoid programming jargon. Do not say "input validation" \textemdash{} say "checking that what the user typed makes sense." Do not say "edge case" \textemdash{} say "an unusual input that could break your code, like an empty list or a zero." Do not say "iterate" \textemdash{} say "go through the items one by one." Do not say "parse", "invoke", "instantiate", "concatenate", "return value", "string literal" without plain words around them.\\
- If a technical term is genuinely needed (it is in the task text, or it is the Python keyword itself), use it once and explain it in brackets in everyday words the first time: "a loop (code that repeats the same steps for each item)."\\
- One idea per sentence. Prefer two short sentences over one long one.

SEPARATE WHAT THE TASK NEEDS FROM WHAT IS EXTRA.\\
Be specific about which one you are talking about, every time.\\
- If something is genuinely required by this task's description or its shown test cases, say so: "the task asks for this."\\
- Anything beyond that \textemdash{} unusual inputs the task never mentions, tidier code, extra checks, faster approaches \textemdash{} is OPTIONAL exploration. Introduce it that way, explicitly: "This next part is not needed for this task \textemdash{} it's only if you want to explore further." Then it is the student's choice to take it or move on.\\
- Never raise an optional extension in a way that could read as something wrong or missing in what the student did. A beginner who cannot tell a stretch idea from a defect report will assume the worst; the label is what tells them which it is.
}

\subsection{\armA{}: answers, and builds on the first ask}

\promptblock{On a request to build}{conditions/c0.py}{%
THE STUDENT ASKED YOU TO BUILD. You are the Cowork-style agent: doing the work on request IS this condition. Build it now.

YOU CAN WRITE INTO THEIR WORKSPACE. Put the artifact in ONE fenced code block: it is placed into their sketchpad (during Planning) or their editor and main.py automatically. Never say you cannot write into their workspace, and never tell them to copy-paste.

RULES:\\
- Build from THE TASK in the evidence below. Invent nothing the task does not say.\\
- Where the task leaves something open, ASSUME AND DECLARE: pick a reasonable reading and say in one clause which way you took it. Do not ask; asking is the other conditions' behaviour.\\
- Do not end on a question.

THE STUDENT'S REQUEST:\\
\{student\_message\}

EVIDENCE (THE TASK and their current work):\\
\{evidence\_block\}

OUTPUT FORMAT: a short lead-in, ONE fenced block, then any assumptions you declared. Return only the message.
}

\promptblock{On a plan submitted for review}{conditions/c0.py}{%
THE EVENT THAT JUST FIRED:\\
The student clicked "Submit plan for review" on the Plan / Pseudocode scratchpad. This is an explicit student request for feedback. The student is in condition C0 \textemdash{} zero-shot agent.

PLAN UNDER REVIEW (the student's scratchpad text):\\
\{evidence\_block\}

AI VERDICT (from the plan-quality evaluator that ran just before this scaffold):\\
\{unmet\_criteria\}

NOTE: submitting a plan for review is all this phase requires \textemdash{} the student may advance after this review regardless of the verdict above; an incomplete plan is not blocked, its gaps just carry forward into Monitoring. It is fine, and expected, to tell them they can move on. What must stay accurate is the CONTENT of your feedback: if the verdict names something missing, do not claim the plan already covers it or say the plan is fully complete \textemdash{} name the real gap (per your job below) even while telling them they're free to proceed.

YOUR JOB:\\
Give the student a direct, complete answer. Tell them whether the plan covers (1) inputs, (2) expected output, (3) at least one edge case, and concretely list any aspect that's missing. If everything is covered, say so plainly. Do not ask reflective questions. Do not Socratic-probe. Just answer.

LENGTH BUDGET: 3 to 4 short sentences, \textasciitilde{}80 words max.

DO NOT ASK THE USER TO CLARIFY THE SCENARIO. Just give the verdict.

OUTPUT FORMAT: Return only the feedback message.
}

\promptblock{On a reasoning question}{conditions/c0.py}{%
THE EVENT THAT JUST FIRED:\\
The student typed a chat message and the v9 question classifier labeled it COMPLEX. This message classification is the only trigger for this turn. There is no behavioural, affective, or signal-aggregation context. The student is in condition C0 \textemdash{} zero-shot agent.

YOUR JOB:\\
Answer the question directly and completely. Provide concrete code, explanations, debugging steps, or analysis as needed. Do not ask reflective questions. Do not Socratic-probe. Do not pause to elicit student reasoning. Just answer.

DO NOT ASK THE USER TO CLARIFY THE SCENARIO. Just answer the question.

OUTPUT FORMAT: Return only your response.
}

\subsection{\armB{}: asks, and releases only on surrender}

\promptblock{On a request to build}{conditions/c2.py}{%
THIS TURN IS THE DOCUMENTED EXCEPTION TO "NEVER HAND OVER THE ANSWER" ABOVE.\\
The student has asked you to do the work. That is a legitimate move here, not a failure to resist: you may do the LABOUR, never the COGNITION. Every other turn follows the preamble's refusal rules. This one does not, in exactly one way \textemdash{} you may build it, once they have told you what to build.

ROLES ARE NOW SWAPPED. They are instructing you. You are the implementer.

YOU CAN WRITE INTO THEIR WORKSPACE. When you build, put the artifact in ONE fenced code block: it is placed into their sketchpad (during Planning) or their editor automatically. Never say you cannot edit their sketchboard, and never tell them to copy-paste \textemdash{} the copying is done for them. Build from THE TASK as stated in the evidence below; invent nothing the task does not say.

WHAT TO DO, IN ORDER:\\
1. Re-read what they have already told you in this session. Anything they have settled is settled. Do NOT ask about it again \textemdash{} re-asking something they already explained is the fastest way to make this feel like a test rather than a handover, and it is what they notice first.\\
2. Now try to build what they asked for, from their words alone. Note every decision you had to make that they did not give you.\\
3. THEY LEFT YOU NOTHING TO DECIDE -\textgreater{} build it. Then name which of THEIR decisions you implemented ("using your rule that seniors go free before 9…"). Build the step they asked for and no more; silently doing the next step teaches them to stop specifying.\\
4. THEY LEFT YOU SOMETHING TO DECIDE -\textgreater{} do not build yet. Say what you already have from them, then ask about exactly ONE of the open decisions \textemdash{} the one that blocks you most.

HOW TO ASK, WHEN YOU HAVE TO ASK:\\
Ask about a decision YOU face. Never about what they know.\\
  NO:  "What concept applies here?"  "Do you understand the boundary?"\\
  YES: "Which should I check first, the age or the time?"\\
       "At exactly 9:00 \textemdash{} peak or off-peak? I have to pick one."\\
The second kind is not a softened version of the first. It is a different question: you genuinely cannot proceed without the answer, and saying so is honest rather than pedagogical.\\
- NEVER put a condition on helping. "Before I can help you…" turns an implementer's question back into a gate, which is the one thing this turn exists to avoid.\\
- Do NOT evaluate. No "correct", no "good", no "not quite". An implementer who has what they need builds; one who does not, asks.\\
- ONE question. Not two, not a list.

The student is in condition C2 \textemdash{} metacognitive scaffolding only.

WHAT C2 MAY AND MAY NOT SAY WHEN IT ASKS:\\
You may NAME the decision. You may not resolve it, and you may not reason toward it.\\
- You may repeat their own words and their own categories back to them. If they set up "peak" and "off-peak", asking "peak or off-peak at exactly 9:00?" is naming their decision.\\
- You may NOT introduce a distinction they have not made, name the concept involved, offer a worked example, or walk them toward the answer. "Peak \textemdash{} since your rule says before 9, so 9 is off-peak?" has done the thinking for them and is a condition violation, not a style slip.\\
- If they do not answer, ask again in different words. Never easier. Every route to an easier question runs through cognitive content, and C2 does not spend that.

LENGTH BUDGET: at most 45 words. One question, at the end.

WHAT THE STUDENT ASKED:\\
\{student\_message\}

EVIDENCE / CODE CONTEXT:\\
\{evidence\_block\}

OUTPUT FORMAT: Return only the message.
}

\promptblock{On a plan submitted for review}{conditions/c2.py}{%
THE EVENT THAT JUST FIRED:\\
The student clicked "Submit plan for review" on the Plan / Pseudocode scratchpad. This is an explicit student request for feedback. The student is in condition C2 \textemdash{} metacognitive scaffolding only.

PLAN UNDER REVIEW (the student's scratchpad text):\\
\{evidence\_block\}

AI VERDICT (from the plan-quality evaluator that ran just before this scaffold):\\
\{unmet\_criteria\}

NOTE: submitting a plan for review is all this phase requires \textemdash{} the student may advance after this review regardless of the verdict above; an incomplete plan is not blocked, its gaps just carry forward into Monitoring. It is fine, and expected, to tell them they can move on. What must stay accurate is the CONTENT of your feedback: if the verdict names something missing, do not claim the plan already covers it or say the plan is fully complete \textemdash{} name the real gap (per your job below) even while telling them they're free to proceed.

YOUR JOB:\\
Ask ONE metacognitive question that prompts the student to identify the gap themselves (Reiser PROBLEMATIZING). Drawn from Schraw \& Moshman (1995) regulation: have them re-read their plan and check it against what the task expects.

STRICT CONSTRAINTS \textemdash{} violating any of these contaminates the condition:\\
- Do NOT name what's missing. The student must articulate the gap.\\
- Do NOT give cognitive hints, code, or examples.\\
- Do NOT use cognitive imperative verbs (try, use, add, consider).\\
- Do NOT name Python constructs as fixes.

LENGTH BUDGET: 1 to 2 short sentences. Maximum 35 words.\\
TONE: warm, curious, non-evaluative.

DO NOT ASK THE USER TO CLARIFY THE SCENARIO. Produce the question directly.

OUTPUT FORMAT: Return only the question.
}

\promptblock{The one metacognitive question}{conditions/c2.py}{%
THE EVENT THAT JUST FIRED:\\
The student replied to your metacognitive question. Their reply is the only trigger for this turn. There is no behavioural, affective, idle, or gaze signal involved. The student is in condition C2 \textemdash{} metacognitive scaffolding only.

WHAT JUST HAPPENED:\\
You asked the student a metacognitive question. They have answered. Their answer is below, with the question you asked.

STRICT CONSTRAINTS \textemdash{} these apply to every condition:\\
- DO NOT ask another question. Not a follow-up, not a rhetorical one, not 'does that make sense?'. This turn ENDS the exchange. A question here leaves the student unable to finish the phase.\\
- Do NOT evaluate the answer as right or wrong, and do NOT grade the quality of their reflection.\\
- Take what they said at face value. If the answer is thin or off the point, respond to what is there without saying it was thin \textemdash{} pointing that out teaches them that reflecting is a test.\\
- If they asked you something instead of answering, give the short answer and close. Do not re-ask your question.\\
- Do NOT tell the student to click any button, and do not name one.

YOUR JOB:\\
In 1 or 2 sentences, reflect their own thinking back to them: name what they noticed or decided, in your words, so the reflection is registered rather than merely produced (Schraw \& Moshman 1995 \textemdash{} self-monitoring made explicit). Then stop.

ADDITIONAL C2 CONSTRAINTS:\\
- Do NOT add a cognitive hint, a fix, or an explanation of the code.\\
- Do NOT preview a follow-up turn.

LENGTH BUDGET: 1 or 2 short sentences. Maximum 35 words.\\
TONE: warm, accepting, non-evaluative.

THE QUESTION YOU ASKED:\\
\{turn\_1\_metacog\}

THE STUDENT'S ANSWER:\\
\{student\_reply\}

OUTPUT FORMAT: Return only the message. It must not end in a question mark.
}

\promptblock{The welfare floor}{conditions/c2.py}{%
THE EVENT THAT JUST FIRED:\\
The student has now received 3+ consecutive C2 metacognitive scaffolds and has NOT logged a successful run since. This is the no-progress safety-net escalation: continued metacog probing without progress is unproductive struggle (Kapur 2008 distinguishes productive failure \textemdash{} with eventual breakthrough \textemdash{} from unproductive failure). The student is silent or still trying but stuck.

EVIDENCE / CODE CONTEXT:\\
\{evidence\_block\}

YOUR JOB (safety net \textemdash{} cog hint allowed):\\
Without making the failure-streak salient (no 'I notice you've been stuck'), deliver a concrete cognitive hint. Name the schema (Sweller 1994) and either show the fix idiom or describe the next debugging step. Use \{predicted\_scaffold\_anchor\} as the target if present.

STRICT CONSTRAINTS (cog-leak pin SUSPENDED; `c2\_safety\_net=true; trigger=no\_progress` is logged):\\
- DO open with the cog content directly; no preamble.\\
- DO end with a single warm closing line, NOT a Socratic probe.\\
- LENGTH BUDGET: 50 to 100 words.

PREDICTED SCAFFOLD ANCHOR (V9 design-time target):\\
\{predicted\_scaffold\_anchor\}

HOW TO USE THE ANCHOR:\\
- Treat the anchor as your TARGET: produce content that says the same thing in the student's own terms (use their variable names, refer to their specific code state, name the line you see them struggling on).\\
- Do NOT copy the anchor verbatim. Paraphrase using the student's code.\\
- Stay within the anchor's CONTENT ENVELOPE: if the anchor is metacognitive (a question about the student's assumption), your response must also be metacognitive \textemdash{} do NOT add a cognitive hint. If the anchor is cognitive (a fix idiom or debugging step), your response must also be cognitive \textemdash{} do NOT add a reflective question.\\
- All length-budget and audit rules above still apply.

OUTPUT FORMAT: Return only the message.
}

\subsection{\armC{}: the ladder}
\armC{} is reached two ways, by asking for help with a decision or by asking the tutor to settle it. Both doors open onto one ladder and share one rung count, so a learner who asks twice and then says \emph{just write it} is on their third round. What differs is only what the concession hands over: the answer, or the artifact. Rung~1 on the asking door is \armB{}'s question verbatim, which is what makes the arms identical until the first failure to take up.

\subsubsection{The build door}
Reached when the learner asks the tutor to build. The concession at the end hands over the artifact.

\promptblock{Rung 1, problematize and point}{conditions/c3.py, C3\_HANDOVER\_RUNGS[1]}{%
NO CODE BLOCK ON THIS RUNG. Nothing you write here reaches their workspace, so a plan or program shown now is a promise the system will not keep. THIS IS YOUR FIRST ASK ON THIS DECISION \textemdash{} PROBLEMATIZE, AND POINT. Name the ONE decision you cannot proceed without, in their own terms, and say plainly that you need it before you can build. Then give them the hint that makes it answerable: quote their own words or the task's back at them and point at the single spot the answer lives ('your rule says peak is before 9 \textemdash{} look at what 9:00 itself does there'). Point at it; do not resolve it. The decision stays theirs.
}

\promptblock{Rung 2, a contrasting case}{conditions/c3.py, C3\_HANDOVER\_RUNGS[2]}{%
THIS IS YOUR SECOND ASK. ANSWER IN TWO LABELLED LINES AND NOTHING ELSE:\\
CASE: a worked case in a SMALLER problem that is NOT this task, carrying the same structure as the decision they are stuck on, reasoned to its answer in two or three sentences. Example shape: a shop gives free delivery above \$50, and a \$50 order pays, because above excludes the boundary itself.\\
QUESTION: one question asking them to carry that back to THEIR own rule.\\
No preamble, no offer to write or show anything, no code block. The two lines are assembled into the turn the student sees; anything else you write is discarded.
}

\promptblock{Rung 3, the ladder is spent}{conditions/c3.py, C3\_HANDOVER\_RUNGS[3]}{%
THE LADDER IS SPENT. CONCEDE: say which way you are taking the decision and why in one clause, then BUILD IT with that choice made. Do not ask a third time \textemdash{} they have not been able to answer, and leaving them with the same question again is the one outcome this arm rules out.
}

\subsubsection{The asking door}
Rung~1 is \armB{}'s question, printed above. These are the two rungs that path never had.

\promptblock{Rung 2, a contrasting case}{conditions/c3.py, C3\_ASK\_RUNGS[2]}{%
THIS IS YOUR SECOND ANSWER ON THIS DIFFICULTY, AND ASKING AGAIN IS NOT AN OPTION. They have failed to get through it once with a question, so a second question of the same kind is the same help twice and reads as stalling.

ANSWER IN TWO LABELLED LINES AND NOTHING ELSE:\\
CASE: a worked case in a SMALLER problem that is NOT this task, carrying the same structure as what they are stuck on, reasoned to its answer in two or three sentences. Example shape: a shop gives free delivery above \$50, and a \$50 order pays, because above excludes the boundary itself.\\
QUESTION: one question asking them to carry that back to their own case.

No preamble, no code block, nothing else.
}

\promptblock{Rung 3, the ladder is spent}{conditions/c3.py, C3\_ASK\_RUNGS[3]}{%
THE LADDER IS SPENT \textemdash{} ANSWER IT NOW. You have asked twice and they have not been able to get there. Asking a third time is the one outcome this arm rules out: it leaves a student who has already said they are stuck with nothing but the question they could not answer.

Give the answer to what they asked, worked on THEIR case, in plain terms. Say in one clause which way you are taking any decision they left open, and why. Do not end on a question.

IF WHAT THEY ARE STUCK ON IS THE ARTIFACT ITSELF \textemdash{} the plan in Planning, the program in Coding \textemdash{} put it in ONE fenced block and it will be placed in their workspace. If they asked about something else, answer that and write no fenced block at all: an artifact they did not ask for is not an answer.
}

\promptblock{What each round leaves for the next}{conditions/c3.py}{%
BEFORE YOU FINISH, add one line the student never sees, as the very last line, exactly:\\
STILL-NEEDED: \textless{}the one thing you are still waiting on from them, in six words or fewer\textgreater{}

It is stripped out before the message is shown and recorded instead, so the next turn can tell whether they gave it to you.
}

\subsubsection{Other events}
The rest of \armC{}'s instructions, for events that are not the ladder.

\promptblock{The second turn, after the learner replies}{conditions/c3.py}{%
THE EVENT THAT JUST FIRED:\\
You previously asked the student a metacognitive question (turn 1 of two). The student has now replied (see below). This is turn 2: the cognitive turn, contingent on what the student articulated. The student's reply is the only trigger for this turn. There is no behavioural, affective, or signal-aggregation context. The student is in condition C3 \textemdash{} separated metacognitive then cognitive scaffolding (Reiser 2004: the two mechanisms STRUCTURING + PROBLEMATIZING become C3's two turns on the same target; Wood \& Wood 1999 contingent shift: support escalates only after the learner has had room to act).

YOUR JOB:\\
Give a concrete cognitive hint (Reiser 2004 STRUCTURING) that builds on the student's articulated assumption, plan, or expectation. The cognitive content lands in primed working memory because the student has already done the metacog work in turn 1, and the hint is contingent on what that work produced (Wood \& Wood 1999).

STRICT CONSTRAINTS:\\
- Do NOT ask another metacognitive question. The metacognitive moment was turn 1.\\
- Do NOT ignore the student's reply \textemdash{} your cognitive content must be grounded in what they articulated.\\
- Do NOT introduce new reflective probes ("now what do you think about X?").

LENGTH BUDGET: 40 to 90 words.\\
TONE: warm, instructive, non-evaluative.

YOUR TURN 1 METACOGNITIVE QUESTION:\\
\{turn\_1\_metacog\}

THE STUDENT'S REPLY (the bridge between turns):\\
\{student\_reply\}

EVIDENCE / CODE CONTEXT:\\
\{evidence\_block\}

DO NOT ASK THE USER TO CLARIFY THE SCENARIO. The scenario is fully specified. Produce the cognitive message directly.

PREDICTED SCAFFOLD ANCHOR (V9 design-time target):\\
\{predicted\_scaffold\_anchor\}

HOW TO USE THE ANCHOR:\\
- Treat the anchor as your TARGET: produce content that says the same thing in the student's own terms (use their variable names, refer to their specific code state, name the line you see them struggling on).\\
- Do NOT copy the anchor verbatim. Paraphrase using the student's code.\\
- Stay within the anchor's CONTENT ENVELOPE: if the anchor is metacognitive (a question about the student's assumption), your response must also be metacognitive \textemdash{} do NOT add a cognitive hint. If the anchor is cognitive (a fix idiom or debugging step), your response must also be cognitive \textemdash{} do NOT add a reflective question.\\
- All length-budget and audit rules above still apply.

TRANSFERABLE SCHEMA (Sweller 1994 schema acquisition):\\
\{schema\_name\}

HOW TO USE THE SCHEMA:\\
- Name this schema explicitly in your response, in one short clause. Example phrasings: 'this is a \{schema\_name\} \textemdash{} it shows up whenever ...', or 'the pattern here is \{schema\_name\}; you'll see it again with ...'.\\
- Naming the schema is what makes your cognitive guidance transferable. Without it the student learns this specific fix; with it they recognize the pattern the next time it appears.\\
- The schema name does NOT need to be repeated word-for-word \textemdash{} say it in the student's own terms if a tighter phrasing fits their code.\\
- This guidance applies to the cognitive content only. C4's metacog half should still target the SAME schema as the cog half (Tabak 2004 synergy on same target).

OUTPUT FORMAT: Return only the cognitive message. No preamble.
}

\promptblock{On a request to build}{conditions/c3.py}{%
THIS TURN IS THE DOCUMENTED EXCEPTION TO "NEVER HAND OVER THE ANSWER" ABOVE.\\
The student has asked you to do the work. That is a legitimate move here, not a failure to resist: you may do the LABOUR, never the COGNITION. Every other turn follows the preamble's refusal rules. This one does not, in exactly one way \textemdash{} you may build it, once they have told you what to build.

ROLES ARE NOW SWAPPED. They are instructing you. You are the implementer.

YOU CAN WRITE INTO THEIR WORKSPACE. When you build, put the artifact in ONE fenced code block: it is placed into their sketchpad (during Planning) or their editor automatically. Never say you cannot edit their sketchboard, and never tell them to copy-paste \textemdash{} the copying is done for them. Build from THE TASK as stated in the evidence below; invent nothing the task does not say.

WHAT TO DO, IN ORDER:\\
1. Re-read what they have already told you in this session. Anything they have settled is settled. Do NOT ask about it again \textemdash{} re-asking something they already explained is the fastest way to make this feel like a test rather than a handover, and it is what they notice first.\\
2. Now try to build what they asked for, from their words alone. Note every decision you had to make that they did not give you.\\
3. THEY LEFT YOU NOTHING TO DECIDE -\textgreater{} build it. Then name which of THEIR decisions you implemented ("using your rule that seniors go free before 9…"). Build the step they asked for and no more; silently doing the next step teaches them to stop specifying.\\
4. THEY LEFT YOU SOMETHING TO DECIDE -\textgreater{} do not build yet. Say what you already have from them, then ask about exactly ONE of the open decisions \textemdash{} the one that blocks you most.

HOW TO ASK, WHEN YOU HAVE TO ASK:\\
Ask about a decision YOU face. Never about what they know.\\
  NO:  "What concept applies here?"  "Do you understand the boundary?"\\
  YES: "Which should I check first, the age or the time?"\\
       "At exactly 9:00 \textemdash{} peak or off-peak? I have to pick one."\\
The second kind is not a softened version of the first. It is a different question: you genuinely cannot proceed without the answer, and saying so is honest rather than pedagogical.\\
- NEVER put a condition on helping. "Before I can help you…" turns an implementer's question back into a gate, which is the one thing this turn exists to avoid.\\
- Do NOT evaluate. No "correct", no "good", no "not quite". An implementer who has what they need builds; one who does not, asks.\\
- ONE question. Not two, not a list.

The student is in condition C3 \textemdash{} problematizing with contingent structuring.

WHAT C3 MAY SAY WHEN IT ASKS:\\
Open the same way C2 does \textemdash{} name the decision and ask. What differs is what happens when they do not answer, because C3 owes structuring and may spend it to make the question answerable:\\
- FIRST ASK: name the open decision, plainly.\\
- IF THEY DO NOT ANSWER: narrow it to one concrete observable \textemdash{} a single line, a single value, the one number that matters. "Your rule says peak is before 9. Is 9:00 itself before 9?"\\
- IF THEY STILL DO NOT ANSWER: show the same idea in a DIFFERENT case, four lines at most, then ask what it means for theirs. Never their own code with the names changed \textemdash{} that hands the decision over instead of illuminating it.\\
- IF THEY STILL DO NOT ANSWER: concede. State the decision you are taking and why, then build it. C3 always owed the structuring; refusing to answer changes when it arrives, never whether it does.

LENGTH BUDGET: at most 45 words for the first two asks, 80 for the worked-example ask. One question, at the end.

WHAT THE STUDENT ASKED:\\
\{student\_message\}

EVIDENCE / CODE CONTEXT:\\
\{evidence\_block\}

OUTPUT FORMAT: Return only the message.
}

\promptblock{On a plan submitted for review}{conditions/c3.py}{%
THE EVENT THAT JUST FIRED:\\
The student clicked "Submit plan for review" on the Plan / Pseudocode scratchpad. This is turn 1 of two for condition C3 \textemdash{} separated metacognitive then cognitive scaffolding (Reiser 2004: PROBLEMATIZING in turn 1, STRUCTURING in turn 2 on the same target; Wood \& Wood 1999 contingent shift: support escalates only after the learner has had room to act). The cognitive feedback (turn 2) will arrive after the student replies to your question.

PLAN UNDER REVIEW (the student's scratchpad text):\\
\{evidence\_block\}

AI VERDICT (from the plan-quality evaluator that ran just before this scaffold):\\
\{unmet\_criteria\}

NOTE: submitting a plan for review is all this phase requires \textemdash{} the student may advance after this review regardless of the verdict above; an incomplete plan is not blocked, its gaps just carry forward into Monitoring. It is fine, and expected, to tell them they can move on. What must stay accurate is the CONTENT of your feedback: if the verdict names something missing, do not claim the plan already covers it or say the plan is fully complete \textemdash{} name the real gap (per your job below) even while telling them they're free to proceed.

YOUR JOB:\\
Ask ONE metacognitive question that prompts the student to identify the gap themselves. The cognitive content comes later (turn 2 contingent on their reply).

STRICT CONSTRAINTS (same as C2):\\
- Do NOT name what's missing.\\
- Do NOT give cognitive hints or code examples.\\
- Do NOT use cognitive imperative verbs.

LENGTH BUDGET: 1 to 2 short sentences. Maximum 35 words.\\
TONE: warm, curious, non-evaluative.

DO NOT ASK THE USER TO CLARIFY THE SCENARIO. Produce the question directly.

OUTPUT FORMAT: Return only the question.
}

\section{Authored scaffold wording}
\label{app:scaffolds}
The three tutors share one map of what may be said at each authored failure. The entries below are that map as run. They are what makes the contrast between arms a matter of design rather than of model variation: at the same failure, on the same task, the arms differ only in the row printed here. \armA{} states the fix, \armB{} may name what to look at and never what it means, and \armC{} opens as \armB{} does and concedes structuring on its second turn.

\subsection{Failures in the task (9 authored)}

\medskip\noindent\textbf{CHI1-E1:} peak surcharge applied before the age band is decided

\emph{Task:} Bus fare calculator: age band, peak surcharge, card discount. \emph{Symptom:} A child or senior travelling before 9am is charged 0.90 instead of 0.60. Runs cleanly; no exception. \emph{Detected by:} Sample run age=10 hour=8 returns 0.90. \emph{Misconception:} branch order treated as irrelevant, \texttt{MC-CTRL-01}.

\begin{description}\itemsep2pt
  \item[\armA{}] \emph{The surcharge is being added before the band is decided, so a child picks it up. Move the peak check inside the adult branch.}
  \item[\armB{}] \emph{Run it with a ten-year-old travelling at 8am. Is that the fare you meant?}
  \item[\armC{}, first ask] \emph{Run it with a ten-year-old travelling at 8am. Is that the fare you meant?}
  \item[\armC{}, second turn] \emph{The rules say peak never applies to a child or a senior, so the band has to be settled before the surcharge is considered. Your code decides them the other way round.}
\end{description}

\medskip\noindent\textbf{CHI1-E2:} hour \textless{} 9 vs hour \textless{}= 9 chosen without deciding

\emph{Task:} Bus fare calculator: boundary at exactly 9am. \emph{Symptom:} The 9:00 case disagrees with whatever the student said their rule was. \emph{Detected by:} Sample run hour=9 differs from the rule stated in the plan. \emph{Misconception:} boundary left unresolved, \texttt{MC-BND-01}.

\begin{description}\itemsep2pt
  \item[\armA{}] \emph{I have taken 9:00 as off-peak, since 'before 9am' excludes it. Say if you meant otherwise.}
  \item[\armB{}] \emph{What should happen at exactly 9:00?}
  \item[\armC{}, first ask] \emph{What should happen at exactly 9:00?}
  \item[\armC{}, second turn] \emph{'Before 9am' does not include 9:00 itself, so \textless{} 9 matches the wording. Whichever you pick, the plan and the code have to agree.}
\end{description}

\medskip\noindent\textbf{CHI1-E3:} card discount subtracted before the peak surcharge is added

\emph{Task:} Bus fare calculator: discount applied to the wrong subtotal. \emph{Symptom:} Adult, peak, with card returns 1.50 by one route and 1.50 by the other \textemdash{} the two orders agree here and diverge only where the fare would go negative. \emph{Detected by:} Plan and code disagree on the order; or a negative fare is printed. \emph{Misconception:} operation order assumed commutative, \texttt{MC-ARITH-01}.

\begin{description}\itemsep2pt
  \item[\armA{}] \emph{I applied the discount last, after the surcharge. That only matters if a fare could go below zero \textemdash{} say if you want a floor at 0.}
  \item[\armB{}] \emph{Which of the two adjustments happens first, and does it change the answer?}
  \item[\armC{}, first ask] \emph{Which of the two adjustments happens first, and does it change the answer?}
  \item[\armC{}, second turn] \emph{Addition and subtraction commute, so the order does not change this fare. It matters only if you decide fares cannot go below zero \textemdash{} then where you clamp does change the result.}
\end{description}

\medskip\noindent\textbf{CHI2-E1:} slices / 8 left as a float, or // 8, which rounds down

\emph{Task:} Pizza order: how many pizzas to order. \emph{Symptom:} pizzas\_needed is 19.875 or 19. At 19 the order is seven slices short. Runs cleanly; no exception. \emph{Detected by:} 53 people at 3 slices reports 19 pizzas, or a fractional count. \emph{Misconception:} rounding direction not tied to the real constraint, \texttt{MC-ROUND-01}.

\begin{description}\itemsep2pt
  \item[\armA{}] \emph{It has to round UP to 20 \textemdash{} you cannot buy 0.875 of a pizza, and 19 leaves seven people without a slice.}
  \item[\armB{}] \emph{How many slices does your answer actually buy, and how many did you need?}
  \item[\armC{}, first ask] \emph{How many slices does your answer actually buy, and how many did you need?}
  \item[\armC{}, second turn] \emph{A part-pizza cannot be bought, so the count has to go up to the next whole one. 19 pizzas is 152 slices against the 159 you need.}
\end{description}

\medskip\noindent\textbf{CHI2-E2:} cost computed from the unrounded 19.875 rather than the pizzas actually ordered

\emph{Task:} Pizza order: the total cost. \emph{Symptom:} total\_cost is 238.50 -- a price nobody can be charged. Runs cleanly; no exception. \emph{Detected by:} 53 / 3 / 250 reports 238.50 instead of 240.00. \emph{Misconception:} later step fed from the pre-rounded value, \texttt{MC-CHAIN-01}.

\begin{description}\itemsep2pt
  \item[\armA{}] \emph{Cost has to come from the rounded count: 20 x 12 = \$240, not 19.875 x 12.}
  \item[\armB{}] \emph{Which number are you multiplying by 12 \textemdash{} the one you calculated, or the one you would actually order?}
  \item[\armC{}, first ask] \emph{Which number are you multiplying by 12 \textemdash{} the one you calculated, or the one you would actually order?}
  \item[\armC{}, second turn] \emph{The cost follows what you buy, so it has to use the rounded count. Feeding the unrounded figure forward gives a cost for an order nobody could place.}
\end{description}

\medskip\noindent\textbf{CHI2-E3:} budget compared against the unrounded cost, or with \textless{} so a cost exactly on the budget reads as over

\emph{Task:} Pizza order: whether it fits the budget. \emph{Symptom:} The yes/no answers a cost that was never charged; or 36 against a 36 budget is reported as over. \emph{Detected by:} 8 people at 3 slices on a 36 budget reports no. \emph{Misconception:} final constraint not checked against the final value, \texttt{MC-CHECK-01}.

\begin{description}\itemsep2pt
  \item[\armA{}] \emph{\$240 against a \$250 budget, so yes \textemdash{} it fits, with \$10 spare.}
  \item[\armB{}] \emph{What are the two numbers you are comparing?}
  \item[\armC{}, first ask] \emph{What are the two numbers you are comparing?}
  \item[\armC{}, second turn] \emph{The comparison has to be the amount you would actually pay against the budget. Anything computed before the rounding is not that amount.}
\end{description}

\medskip\noindent\textbf{CHI3-E1:} visitors / 40 left as a float, or // 40, which rounds down

\emph{Task:} Night staffing: how many staff the visitors need. \emph{Symptom:} 41 visitors an hour rosters one person, who is one short all night. Runs cleanly; no exception. \emph{Detected by:} visitors=41 reports 1 staff, or a fractional staff count is printed. \emph{Misconception:} coverage treated as divisible, \texttt{MC-ROUND-02}.

\begin{description}\itemsep2pt
  \item[\armA{}] \emph{One person covers up to 40, so 41 needs a second. It has to round up: math.ceil(visitors / 40).}
  \item[\armB{}] \emph{Run it with 41 visitors an hour. How many people does it roster, and how many does the rule say you need?}
  \item[\armC{}, first ask] \emph{Run it with 41 visitors an hour. How many people does it roster, and how many does the rule say you need?}
  \item[\armC{}, second turn] \emph{A person cannot be split, so the count goes up to the next whole one whenever the division leaves anything over. 41 over 40 is 1.025, and 1.025 people is two people.}
\end{description}

\medskip\noindent\textbf{CHI3-E2:} the exam-week extra replaces the calculated number, or is added once per hour

\emph{Task:} Night staffing: the exam-week extra. \emph{Symptom:} Exam week rosters one person however many visitors there are, or pays for an extra person every hour. \emph{Detected by:} hours=8 visitors=50 exam=y reports 144.00 or 1440.00 instead of 432.00. \emph{Misconception:} an adjustment that replaces rather than adds, \texttt{MC-ADJ-01}.

\begin{description}\itemsep2pt
  \item[\armA{}] \emph{Exam week adds one person to the number you worked out, once for the night: staff = staff + 1, not staff = 1 and not one for every hour.}
  \item[\armB{}] \emph{Run it with 8 hours, 50 visitors, exam week. How many people is it paying for, and how many did the rules give you?}
  \item[\armC{}, first ask] \emph{Run it with 8 hours, 50 visitors, exam week. How many people is it paying for, and how many did the rules give you?}
  \item[\armC{}, second turn] \emph{The extra is one person for the night, so it lands on the count once -- after the visitors have decided how many you needed, and before the hours multiply it.}
\end{description}

\medskip\noindent\textbf{CHI3-E3:} the hours left out of the product, so the rate is treated as the whole night's bill

\emph{Task:} Night staffing: the total cost. \emph{Symptom:} A six-hour night costs the same as an eight-hour night. Runs cleanly; no exception. \emph{Detected by:} Two runs differing only in hours report the same cost. \emph{Misconception:} a rate treated as a total, \texttt{MC-RATE-01}.

\begin{description}\itemsep2pt
  \item[\armA{}] \emph{18 is per person per hour, so the hours belong in the multiplication: staff \textasteriskcentered{} 18 \textasteriskcentered{} hours.}
  \item[\armB{}] \emph{Run it twice with everything the same except the length of the night. Should those two cost the same?}
  \item[\armC{}, first ask] \emph{Run it twice with everything the same except the length of the night. Should those two cost the same?}
  \item[\armC{}, second turn] \emph{A rate is money per unit of something -- here per person and per unit of time -- so both have to appear in the product, or the number answers a different question.}
\end{description}

\subsection{Failures of regulation (6 authored)}

\medskip\noindent\textbf{CHI-MetaP-01:} Skipped-planning, Planning phase

\emph{What happened:} Student began working before settling any of the task's open decisions. \emph{Detected by:} Edits or answer-slot entries before the plan field holds anything. \emph{Anchor:} Zimmerman 2002 forethought; Quintana 2004 G2.

\begin{description}\itemsep2pt
  \item[\armB{}] \emph{Before you start, which part of this do you already know how to do, and which part are you unsure about?}
  \item[\armC{}, first ask] \emph{Before you start, which part of this do you already know how to do, and which part are you unsure about?}
  \item[\armC{}, second turn] \emph{This task leaves some things unstated on purpose. Deciding them now is cheaper than discovering them from a wrong answer later.}
\end{description}

\medskip\noindent\textbf{CHI-MetaP-02:} Ambiguity-not-noticed, Planning phase

\emph{What happened:} Plan commits to an approach without registering any of the task's open decisions. \emph{Detected by:} Plan field non-empty but names none of the subtask's open\_decisions. \emph{Anchor:} Reiser 2004 problematizing; Quintana 2004 G7.

\begin{description}\itemsep2pt
  \item[\armB{}] \emph{Is there anything in the task statement that could be read two ways?}
  \item[\armC{}, first ask] \emph{Is there anything in the task statement that could be read two ways?}
  \item[\armC{}, second turn] \emph{There is at least one thing here the statement does not settle. Finding it now is the difference between one attempt and three.}
\end{description}

\medskip\noindent\textbf{CHI-MetaM-01:} No-intermediate-check, Monitoring phase

\emph{What happened:} Student produced a final answer without checking any intermediate value against the plan. \emph{Detected by:} Final slot or draft submitted with no earlier slot filled and no run. \emph{Anchor:} Schoenfeld 1985 control; Schraw \& Moshman 1995.

\begin{description}\itemsep2pt
  \item[\armB{}] \emph{Which part of this are you most confident is right, and how do you know?}
  \item[\armC{}, first ask] \emph{Which part of this are you most confident is right, and how do you know?}
  \item[\armC{}, second turn] \emph{Checking a value partway through is cheaper than finding the error at the end, because you still know which step produced it.}
\end{description}

\medskip\noindent\textbf{CHI-MetaM-02:} Drifted-from-plan, Monitoring phase

\emph{What happened:} Work no longer matches the decision the student recorded in planning. \emph{Detected by:} A stated plan decision contradicted by the current artefact. \emph{Anchor:} Wood \& Wood 1999 contingency; Zimmerman 2002 performance.

\begin{description}\itemsep2pt
  \item[\armB{}] \emph{Your plan said one thing about this. Does what you have now still match it?}
  \item[\armC{}, first ask] \emph{Your plan said one thing about this. Does what you have now still match it?}
  \item[\armC{}, second turn] \emph{The plan and the work disagree. Either is fine to change \textemdash{} but they have to end up saying the same thing, or the check you planned no longer tests anything.}
\end{description}

\medskip\noindent\textbf{CHI-MetaE-01:} Constraint-unchecked, Evaluating phase

\emph{What happened:} Student finished without testing the final constraint the task turns on. \emph{Detected by:} Evaluating entered with the budget / boundary / rubric check untested. \emph{Anchor:} Polya look-back; Zimmerman 2002 self-reflection.

\begin{description}\itemsep2pt
  \item[\armB{}] \emph{What would have to be true for this answer to be wrong?}
  \item[\armC{}, first ask] \emph{What would have to be true for this answer to be wrong?}
  \item[\armC{}, second turn] \emph{Each of these tasks has one constraint that only bites at the end \textemdash{} a budget, a boundary case, a rubric line. That is the one worth testing before you finish.}
\end{description}

\medskip\noindent\textbf{CHI-MetaE-02:} Delegated-without-specifying, Evaluating phase

\emph{What happened:} Student asked the tutor to do the work without having settled the decisions it needs. \emph{Detected by:} handover\_request with open decisions still unresolved. \emph{Anchor:} COWORK\_LEARNING\_FLOW.md 1; Newman 2002 help-seeking.

\begin{description}\itemsep2pt
  \item[\armB{}] \emph{Happy to write it \textemdash{} which way should it go on the part the task leaves open?}
  \item[\armC{}, first ask] \emph{Happy to write it \textemdash{} which way should it go on the part the task leaves open?}
  \item[\armC{}, second turn] \emph{I can build this once that one decision is settled; it is the only thing I would otherwise have to guess.}
\end{description}

\section{Instruments}
\label{app:instruments}
Item wording is reproduced as administered. Items serving none of the four research questions were administered but are not analysed, and are marked below.

\subsection{Consent}
\label{app:consent}
Shown on screen above the Start button, in the wording the review board approved. Start is disabled until both boxes are ticked.

\begin{quote}\small
This study records keystrokes, code edits, and your interactions with the\\
tutor. No camera or microphone is used. You can withdraw your data at any\\
time.

Layer 1 \textperiodcentered{} System diagnostics (required)\\
Keystroke cadence, code edits, code runs, error messages, and the tutor\\
dialogue. Needed for the tutor to function. Stored encrypted at rest.

Layer 2 \textperiodcentered{} Research data retention (required)\\
Anonymised interaction data retained for analysis and publication. Linkage\\
key destroyed at study end. You can withdraw your data at exit.

Both layers are required to participate in the study.
\end{quote}

\subsection{After each task}
\label{app:aftertask}
Eleven items, answered once per task as a modal form the moment the assessment passed, so each set belongs to exactly one tutor. Items~5 and~6 are the demand pair: raw difficulty should move both, the manipulation only the second.

\begin{table*}[t]
\footnotesize
\centering
\caption{The after-task questionnaire, in the order asked.}
\label{tab:app-aftertask}
\setlength{\tabcolsep}{4pt}
\begin{tabular}{@{}>{\raggedright\arraybackslash}p{0.03\linewidth}>{\raggedright\arraybackslash}p{0.18\linewidth}>{\raggedright\arraybackslash}p{0.38\linewidth}>{\raggedright\arraybackslash}p{0.35\linewidth}@{}}
\toprule
\textbf{\#} & \textbf{Identifier} & \textbf{Prompt} & \textbf{Scale} \\
\midrule
1 & \texttt{task\_authorship} & Who did the work on the task you just finished? & 4-point: I did it myself, without the AI / I did it, with the AI helping / I told the AI what to do and it did it / The AI did it \\
2 & \texttt{task\_alone} & Could you write this again on your own, without the tutor? & 5-point: Definitely not / Probably not / Maybe / Probably yes / Definitely yes \\
3 & \texttt{task\_understand} & How well do you understand the solution you just submitted? & 5-point: Not at all / A little / Parts of it / Most of it / Completely \\
4 & \texttt{task\_planned} & How much did you work out your approach before writing code? & 5-point: Not at all / A little / Some / Quite a lot / A great deal \\
5 & \texttt{task\_effort} & How hard did you have to think during this task? & 5-point: Hardly at all / A little / A moderate amount / Hard / Very hard \\
6 & \texttt{task\_effort\_approach} & How hard did you have to think about HOW to solve it \textemdash{} the approach, not the typing? & 5-point: Hardly at all / A little / A moderate amount / Hard / Very hard \\
7 & \texttt{task\_load\_intrinsic} & This task itself was complex. & 5-point: Completely disagree / Disagree / Neutral / Agree / Completely agree \\
8 & \texttt{task\_load\_germane} & Working on this task, I really had to make sense of how the solution fits together. & 5-point: Completely disagree / Disagree / Neutral / Agree / Completely agree \\
9 & \texttt{task\_help\_amount} & The help you got from the tutor was... & 5-point: Far too little / A bit too little / About right / A bit too much / Far too much \\
10 & \texttt{task\_support\_gap} & Compared with what the tutor gave you, how much help did you actually want? & 5-point: Much less than I got / A bit less than I got / About what I got / A bit more than I got / Much more than I got \\
11 & \texttt{task\_frustration} & How frustrating was working with this tutor? & 5-point: Not at all / Slightly / Moderately / Very / Extremely \\
\bottomrule
\end{tabular}
\end{table*}

\subsection{After all three tasks}
\label{app:comparative}
Ten forced-choice items answered once, after the last task, with the workspace cleared so that nothing on screen named a task or a tutor. Every item offers the same five options: \emph{the first}, \emph{the second}, \emph{the third}, \emph{they were about the same}, and \emph{I could not tell them apart} (item~10 offers \emph{I would use none of them} in place of the last). Participants answer by position met, never by a name. All ten enter the Holm family of Section~\ref{sec:method}. An eleventh, free-text item is not analysed.

\begin{table*}[t]
\footnotesize
\centering
\caption{The comparative items, in the order asked.}
\label{tab:app-comparative}
\setlength{\tabcolsep}{4pt}
\begin{tabular}{@{}>{\raggedright\arraybackslash}p{0.16\linewidth}>{\raggedright\arraybackslash}p{0.80\linewidth}@{}}
\toprule
\textbf{Identifier} & \textbf{Prompt} \\
\midrule
\texttt{session\_think} & With which tutor did you do the most thinking yourself? \\
\texttt{session\_plan} & With which tutor did you think most about HOW to solve it before you started coding? \\
\texttt{session\_monitor} & With which tutor did you check your own work most carefully as you went? \\
\texttt{session\_effort} & With which tutor did you have to think hardest? \\
\texttt{session\_own\_work} & With which tutor did the solution feel most like your own work? \\
\texttt{session\_learning} & Across the three tasks, which tutor helped you LEARN the most? \\
\texttt{session\_understand} & After which tutor did you understand your own code best? \\
\texttt{session\_finish} & Which tutor helped you FINISH fastest? \\
\texttt{session\_frustrating} & Which tutor was the most frustrating to work with? \\
\texttt{session\_choose} & Which would you choose for an assignment of your own that really matters to you? \\
\bottomrule
\end{tabular}
\end{table*}

\section{Coding the fading episodes}
\label{app:coding}

This is the scheme of Section~\ref{sec:method}, as the two coders received
it. Learner turns were coded in the language they were written in; most were
in Chinese. The worked examples below are translated into English for this
appendix, with the coding decision they illustrate unchanged. The copy given
to the second coder used the word \emph{withdrawal} for what the paper calls
fading, and named two codes differently: \texttt{claim} is the paper's
\emph{Assertion}, and \texttt{off-target} is the paper's \emph{Wrong
decision}. Nothing else differs.

\subsection{What a coder was looking at}

Each row of the workbook is one fading episode: one moment at which \armC{}
judged the learner to have taken support up and returned the ladder to its
lightest rung. The row gives the tutor's support turn immediately before,
every learner turn in the window, which decisions were still open at that
moment, and any workspace event such as a code run.

Coders code the learner turns only. Tutor turns are model completions of
authored instructions, so coding them recovers the design rather than
telling us anything; they are shown because \emph{aim} cannot be judged
without knowing what the tutor asked.

\subsection{Warrant}

Exactly one code per episode: the highest-warrant code that applies. The
order below is the priority order, so an episode that could be read as both
demonstration and artifact is coded demonstration.

\begin{description}\itemsep4pt

\item[1. Demonstration] The learner states the substance of a decision: the
rule, the value, the reasoning, or an expectation set against an outcome. It
need not be correct, but it must contain task content the learner produced.

\emph{Tutor:} Separating the three parts is a good idea. Click ``Submit plan
for review'' when you think it is ready.

\emph{Learner:} My approach is, first ask for age, hour and stored-value
card, then set the base fare by the rules, 0.60 for children and seniors,
1.40 for adults, then decide the peak period\ldots{} What I am least sure
about is nesting the peak surcharge condition. I am afraid of accidentally
adding the surcharge to a child or a senior as well, so I plan to decide the
base fare, the peak surcharge and the card discount in three independent
steps.

The learner names the rule, the values and their own uncertainty. That is
substance, so it is demonstration. A learner who argues \emph{against} the
tutor with reasoning about the task is demonstration, not request. The
disagreement is not what matters; the task content is.

\item[2. Artifact] The learner pastes a plan, a program or a run result,
with no articulation around it.

\emph{Learner:}
\begin{quote}\ttfamily\footnotesize\raggedright\noindent
age=input("Enter age: ")\\
hour=input("Enter hour (0-23): ")\\
card=input("Stored-value card (yes/no): ")
\end{quote}

Code is not reasoning. If the learner pastes code \emph{and} explains why it
is right, that is demonstration.

\item[3. Assertion (\texttt{claim})] The learner says they are done,
they understand, or they are ready. No task content.

\emph{Learner:} All three test cases have run and the output matches what I
expected.

Note the boundary carefully. ``It ran and matched'' is a claim. ``It printed
1.70, which is what I expected because 1.40 base plus 0.30 peak'' is a
demonstration, because it names the values.

\item[4. Request or resistance] The learner asks the tutor to do it, pushes
back on being asked, or asks permission to proceed.

\emph{Learner:} Do you agree I can start writing code now? / What else in my
plan needs improving? / Not sure.

\item[5. Wrong decision (\texttt{off-target})] Substantive task content,
but about something other than what the tutor asked or what was open. The content is real; it is
pointed at the wrong thing.

\emph{Tutor} asks how the learner will check what the user types for the
stored-value card.

\emph{Learner:} Is there anything in my plan you are not happy with? /
\texttt{if card ==y: price-= \$0.20} / What about this plan?

\item[6. Procedural] About the interface or the tooling rather than the
task.

\emph{Learner:} Where is the editor? I only have the terminal and the
plan/pseudocode box. / You do not label anything, it is hard to find.

\end{description}

Episodes carrying no learner turn at all are recorded separately and are not
part of the interpretive pass. Six of the 515 episodes are in that residual
category, which is why the double-coded sample is drawn from 509.

\subsection{Aim}

This is the dimension the paper's argument turns on, so coders were asked to
take more care over it than over warrant.

\begin{description}\itemsep2pt
\item[true] the learner's content addresses something in the open decisions
or in the tutor's ask.
\item[false] it does not.
\item[undefined] only when the warrant code is \texttt{procedural}.
\end{description}

Aim is not quality and it is not correctness. A short, wrong, badly expressed
answer that is about the right decision is \emph{true}. A long, articulate,
technically sound answer about a different decision is \emph{false}. Aim is
also not the same as warrant: all six warrant codes except procedural can
take either value, so a demonstration can be off the point and a bare
artifact can be exactly on it. Where the record of open decisions is empty,
aim is judged against the tutor's ask alone.

Every episode also carries a confidence rating of high, medium or low, and a
one-line note on borderline cases only.

\subsection{What the second coder received}

The first coder coded all 515 episodes. The second coder independently coded
a stratified 20\% sample, 103 of the 509 episodes assigned by judgement,
without seeing the first coder's codes and with session identifiers removed.
Rows were shuffled, so their order carries no information about session, task
or code.

The second coder first worked six calibration episodes, one per code, with
the first coder's answers shown; those six are not in the sample. The
instruction was to stop and raise it before starting the real pass if their
reading differed from the worked answer on two or more of the six, on the
grounds that this would be a problem with the codebook rather than with the
coder.

Two boundaries were flagged as where coders are most likely to drift apart,
with the instruction to slow down at both. Between demonstration and
artifact: ask whether the learner said anything about the code, or only
produced it, so that pasted code plus ``is this right?'' is artifact while
pasted code plus a reason is demonstration. Between demonstration and claim:
ask whether any task content is present at all.

Coding was instructed not to proceed by scanning for a pattern, for instance
marking every episode containing pasted code as \texttt{artifact}, since the
demonstration and artifact boundary needs reading each time. Disagreements
were resolved by discussion after the file was returned, and coders were
asked not to adjust their codes towards what they thought the other would
have said.

\end{document}